\documentclass[10pt]{iopart}
\expandafter\let\csname equation*\endcsname\relax
\expandafter\let\csname endequation*\endcsname\relax
\usepackage{mathtools}
\usepackage{amstext}
\usepackage{amsfonts}
\usepackage{booktabs}
\usepackage{graphicx}
\usepackage[pdfborder={0 0 0.5}]{hyperref}
\usepackage{placeins}
\usepackage{url}
\usepackage{microtype}

\usepackage[backend=biber, sorting=none, bibstyle=ieee, citestyle=numeric-comp]{biblatex}
\newcommand{\iotabar}{\ensuremath{\iota\hspace{-0.45em}\rotatebox{15}{--}}}

\usepackage{xcolor}
\usepackage{soul}
\graphicspath{{./figs/}}

\begin{document}

\title[]{Characterization of NBI-driven shear Alfvén waves in the TJ-II stellarator using Mirnov probes and electrostatic potential fluctuation measurements}
\author{
	P. Pons-Villalonga\(^{1}\),
	Á. Cappa\(^{1}\),
	E. Ascasíbar\(^{1}\),
	O.S. Kozachok\(^{2}\),
	M.B. Dreval\(^{2}\),
	K.J. McCarthy\(^{1}\),
	J. de la Riva Villén\(^{1}\),
	J. Martínez-Fernández\(^{1}\),
	and TJ-II Team
}
\address{\(^1\) Laboratorio Nacional de Fusión, CIEMAT, 28040 Madrid, Spain}
\address{\(^2\) IPP NSC Kharkov Institute of Physics and Technology, Kharkov, Ukraine}
\ead{pedro.pons@ciemat.es}
\ead{alvaro.cappa@ciemat.es}
\date{\today}

\begin{abstract}
	We present the first experimental measurements of the toroidal mode number of shear Alfvén waves in the TJ-II stellarator.
	A series of experiments were carried out in three different magnetic configurations to investigate counter-NBI driven modes.
	Co- and counter- electron-cyclotron current drive was used to modify the rotational transform (\iotabar) profile leading to the destabilization of a varied set of Alfvén eigenmodes with different frequencies and mode numbers.
	To characterize the spatial structure of the modes we have used two Mirnov probe arrays, one dedicated to the measurement of the poloidal mode number and the other, a recently commissioned helical tri-axial array, dedicated to the measurement of the toroidal mode number.
	A heavy ion beam probe, operated in radial sweep mode, was employed to characterize the radial location of the modes.
	We show that the induced changes in \iotabar, that are fundamental when it comes to validation studies, cannot be measured experimentally with motional Stark effect so, instead, the shielding current diffusion equation is solved in cylindrical geometry to estimate these changes.
	We calculate the incompressible shear Alfvén continuum for selected cases using \texttt{STELLGAP} and find reasonable consistency with observations.
	A database with the observed modes has been created, so that it can be used in future work for theory validation purposes.
\end{abstract}

\ioptwocol

\section{Introduction}
\label{sec:introduction}

The validation against experimental results of Alfvén Eigenmodes (AEs) physics models and of the numerical tools that implement these models is indispensable for the informed design of future reactors \cite{fasoliChapter52007}.
The experimental determination of the spatial structure of AEs through the identification of its poloidal (m) and toroidal (n) mode numbers and the measurements of the radial profiles of the electric potential is an integral part of this task.

The excitation of fast-particle induced Alfvén Eigenmodes has been routinely observed in TJ-II \cite{cappaStabilityAnalysis2021, eliseevAlfvenStudyHIBP2022, melnikovDetectionInvestigation2018, nagaokaMitigationNBIdriven2013}.
NBI-driven modes measured in TJ-II plasmas appear as intense, narrow frequency bands in the spectrogram of magnetic, plasma potential or radiation fluctuations allowing, in principle, to have a clear diagnosis of the properties of the modes.

The main objective of the experiments described in this paper was to excite AEs in different scenarios and study their spatial structure, focusing on the mode number identification (\(n, m\)) by means of poloidal and helical arrangements of Mirnov coils \cite{ascasibarMeasurementsMagnetic2022} on the one hand and, on the other hand, on the measurement of the radial profile of perturbed potential using Heavy Ion Beam Probes (HIBP).
The measurement of the poloidal mode structure is routinely done since the poloidal array of Mirnov coils was commissioned and the determination of the poloidal number has been employed in previous studies \cite{jimenez-gomezAlfvenEigenmodes2011,vanmilligenMHDMode2011}.
Only very recently we have added the possibility to determine the toroidal mode number \cite{ascasibarMeasurementsMagnetic2022} and the polarization of the wave detected at the coils locations.
Beyond the interest related to model validation, the difficulty of this measurement, in particular in non-axisymmetric configurations, calls for a complete experimental study where we know that we are varying the spectrum of possible modes and can thus confirm that these variations are reflected in the helical array data and the toroidal mode numbers reconstructed from this data.
Both measurements, magnetics and HIBP, along with the recent development of a synthetic diagnostic for Mirnov coil arrays \cite{pons-villalongaExploringOperational2024} allow us to establish appropriate criteria to support the validity of the mode analysis results.

On-axis Electron Cyclotron Current Drive (ECCD) was chosen as actuator \cite{garciamunozActuators2019} to modify the spectrum of AEs due to its minimal impact on the plasma density and temperature profiles compared to the effect it has on the plasma current, which isolates the rotational transform profile (\iotabar) as the driver of changes in the shear Alfvén wave (SAW) spectrum \cite{cappaStabilityAnalysis2021}.
To obtain a comprehensive database of different excited AEs, we conducted a parameter scan on magnetic configuration and EC beam injection angle.
An essential piece of information when performing validation tasks is the rotational transform profile.
Unfortunately, as we will show here, the changes induced by ECCD or Neutral Beam Current Drive (NBCD), which are important from the point of view of their impact on the Alfvén continuum, are not intense enough to be measurable by \iotabar\ diagnostics such as Motional Stark Effect (MSE) \cite{mccarthySpectrallyResolved2015}.
Therefore we must rely on estimates of the different external current sources and simulate the time evolution of rotational transform by calculating the radial diffusion of the shielding current.
Previous validation studies \cite{ortizHAE2025, ghiozziAEs2024, cappaStabilityAnalysis2021, eliseevGAE2021} carried out with the \texttt{STELLGAP} \cite{spongShearAlfven2003} and \texttt{FAR3d} \cite{varelaStabilityOptimization2024} codes showed predictions consistent with observations.
In these works, no shielding current diffusion was calculated at any moment.
The model used to provide an estimate of the time evolution of the rotational transform depended only on the total value of the toroidal plasma current; a given combined current density profile for the source terms (NBCD, ECCD or OH-induced current) was assumed in \cite{eliseevGAE2021, ghiozziAEs2024} while numerical estimates of the NBCD, ECCD and bootstrap current sources were taken in \cite{ortizHAE2025, cappaStabilityAnalysis2021} to model the final phase of the discharge.

In addition to that, in almost all of these studies, the source of fast ions was uniquely provided by the NBI co-injector (NBI1) while, in general, counter injection with NBI2, less explored experimentally \cite{melnikovAEsProperties2012, melnikovDetectionInvestigation2018}, produces a richer and much more complex spectrum of modes.
Since NBI1 driven AEs have been already the subject of research in the papers we have cited above, the present study focuses only on the analysis of the modes produced by the NBI2 injector in three different magnetic configurations.

This paper is organized as follows: section \ref{sec:exp_setup} contains a description of the experimental set-up and the relevant plasma parameters that were achieved in the experiments.
Then, in section \ref{sec:mag_fluctuations} we describe the mode analysis technique and the whole set of measured mode numbers.
Next, in section \ref{sec:hibp}, after a brief description of the HIBP measurements, we present several examples of the perturbed electrostatic potential profiles associated with the modes.
In section \ref{sec:iotaevol}, we solve the shielding current radial diffusion equation in cylindrical geometry taking the total experimental current as boundary condition and approximate profiles for the external current sources.
This allows us to estimate the evolution of the rotational transform profile.
A short discussion on the difficulty of resolving changes in \iotabar\ using MSE measurements is also included in this section.
Although the main purpose of the paper is the experimental characterization of instabilities in different scenarios, in section \ref{sec:discuss} we will take a couple of cases and discuss their corresponding Alvén continuum calculated with \texttt{STELLGAP}.
Finally, we come to our conclusions and lay the future groundwork for the implementation of the validation studies.

\section{Experimental arrangement and results}
\label{sec:exp_setup}

\begin{figure}[h] \centering
	\includegraphics[width=0.9\columnwidth]{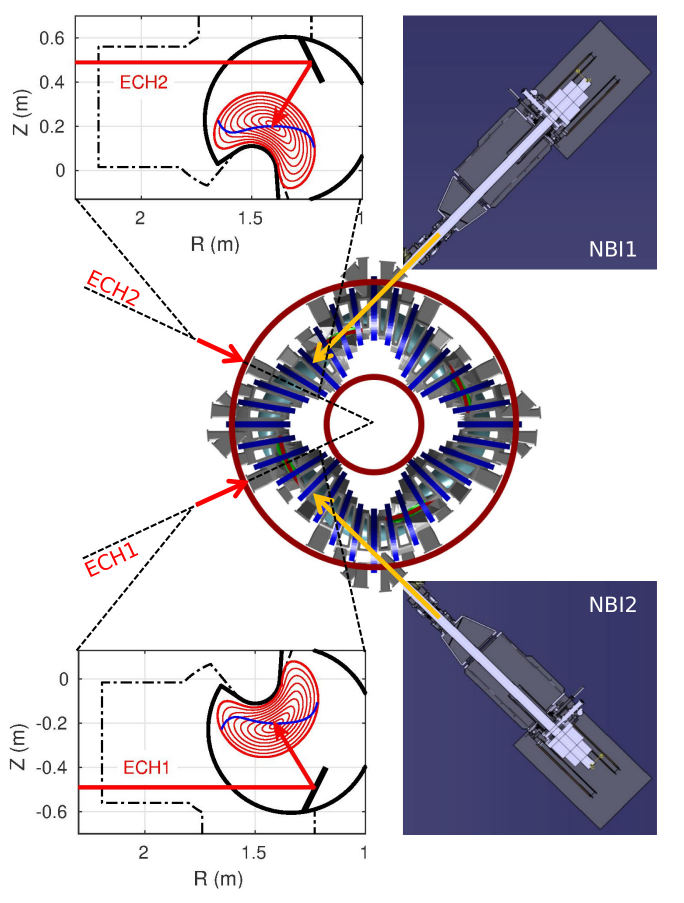}
	\caption{TJ-II stellarator heating systems.
		NBI1 injects tangentially in the direction of the magnetic field (co-injection), while NBI2 does that in the opposite direction (counter-injection).
		ECRH beams (second harmonic X-mode) are launched at stellarator symmetric positions.}
	\label{fig:HeatingSYS}
\end{figure}

TJ-II \cite{ascasibarOverviewTJII2001} is a four period ($N_{fp}=4$) medium-sized stellarator of the heliac type with major radius \(R = 1.5\) m, minor average radius \(a \leq 0.22\) m, magnetic field on axis \(B_0 = 0.95\) T and plasma volume \(V \leq 1\) m\(^{3}\).
NBI sources produce co and counter \(H_0\) tangential beams at 32 kV. Figure \ref{fig:HeatingSYS} shows the launching configuration around the device of both neutral beams and the two electron cyclotron heating beams (second harmonic X-mode).

\begin{figure}[h] \centering
	\includegraphics[]{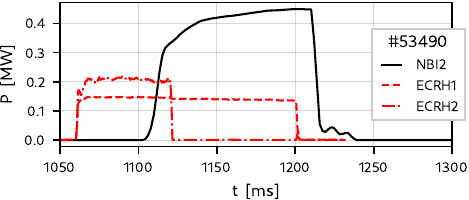}
	\caption{Heating sequence used for all experiments.
		Time traces of ECRH (red) and NBI2 power (black) are shown.}
	\label{fig:heating_scheme}
\end{figure}

To investigate the spatial structure of the AEs destabilized in different magnetic configurations and with variable on-axis ECCD settings, the following experimental set-up was used.
The heating sequence is shown in figure \ref{fig:heating_scheme}.
The plasma is started using ECRH, with both beams switched-on simultaneously.
One of them is set to inject perpendicularly to the magnetic axis (ECRH2) and the other (ECRH1) is launched with different incidence angles that are varied on a shot to shot basis.
After \(\sim 50\) ms, NBI2 is switched on, creating a fast-particle population that heats the plasma when slowing down and at the same time induces current (NBCD) and drives the AEs.
The ECRH2 beam with perpendicular incidence is turned off shortly after the injection of NBI2, while the ECRH1 beam remains on to both drive ECCD and help to maintain a constant plasma density during the rest of the shot.

We have performed the same scan using three distinct magnetic configurations with different values of rotational transform on-axis and very similar low shear.
For our present purposes, although the central rotational transform can be varied between larger margins \(0.9\) and \(2.5\), we will refer to them as low, medium and high \iotabar\ configurations in the text.
Table \ref{tab:confs} shows the central values of \iotabar\ and the correspondence to the labeling used in all figures, which is based on the currents flowing in the main field conductors.
\begin{table}[h] \centering
	\begin{tabular}{@{}lrrr@{}}
		\toprule
		\iotabar    & low         & medium      & high        \\ \midrule
		conf. label & 100\_36\_62 & 100\_44\_64 & 100\_49\_65 \\ \midrule
		\iotabar(0) & 1.47        & 1.55        & 1.6         \\ \bottomrule
	\end{tabular}
	\caption{Equivalence between \iotabar and the current based labeling.}
	\label{tab:confs}
\end{table}

\begin{figure}[h] \centering
	\includegraphics[]{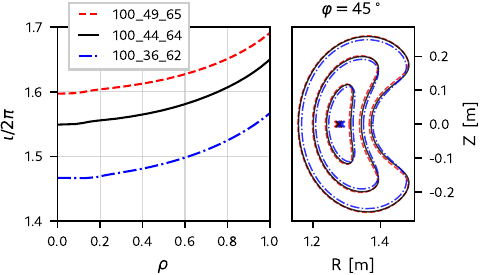}
	\caption{Left: Vacuum rotational transform profiles for the high, medium, and low iota configuration. Right: Comparison of vacuum flux surfaces for the three explored magnetic configurations.}
	\label{fig:iota}
\end{figure}

The vacuum \iotabar\ profiles are represented in figure \ref{fig:iota}, along with a poloidal cross section of the magnetic surfaces of each configuration at $\varphi=45^\circ$.

Exploring configurations with large or lower values of \iotabar\ is generally more demanding in terms of plasma performance and density control. In this case, except for the high \iotabar\ configuration that needed $P_{\text{ECRH1}}=200$ kW to improve density control, the rest of shots were achieved with $P_{\text{ECRH2}}=200$ kW, $P_{\text{ECRH1}}=150$ kW and $P_{\text{NBI2}}=450$ kW.
\begin{table}[h] \centering
	\begin{tabular}{@{}lrrrrr@{}}
		\toprule
		\(n_\parallel\)         & -0.2  & -0.1 & 0  & 0.1  & 0.2  \\ \midrule
		\(\alpha\) [\(^\circ\)] & 101.5 & 95.7 & 90 & 84.3 & 78.5 \\ \bottomrule
	\end{tabular}
	\caption{Equivalence between \(n_\parallel\) and ECRH1 beam injection angle.}
	\label{tab:n_parallel_angle}
\end{table}
\begin{figure}[h] \centering
	\includegraphics[width=0.45\textwidth]{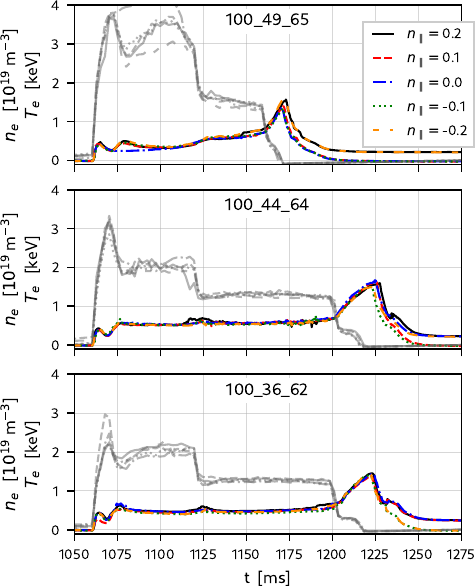}
	\caption{
		Time traces of the electron line densities (colored) and central electron temperature (gray) obtained in the ECCD scan for the three explored magnetic configurations.
	}
	\label{fig:densities}
\end{figure}
As was mentioned in the introduction, the advantage of using on-axis ECCD to modify the spectrum of shear Alfvén waves is that the small changes induced in the plasma density and temperature profiles when steering the beam along the axis have a smaller incidence on the SAW spectrum than the changes produced by variations in the toroidal plasma current.
Five positions of the ECRH1 beam were used to drive different amounts of toroidal plasma current.
They are labeled according to the on-axis wave parallel refraction index in the plasma disregarding refraction effects, \(n_{\parallel} = \cos \alpha\), where \(\alpha\) is the angle between the wave vector of the injected wave and the direction of the magnetic field along the axis. Negative values of \(n_\parallel\) correspond to co-ECCD, generating positive current (increasing \iotabar), while positive ones stand for counter-ECCD, that generates negative current (decreasing \iotabar).
Table \ref{tab:n_parallel_angle} shows the ECRH1 injection angle in degrees for each value of \(n_\parallel\).

\begin{figure}[h] \centering
	\includegraphics[width=0.45\textwidth]{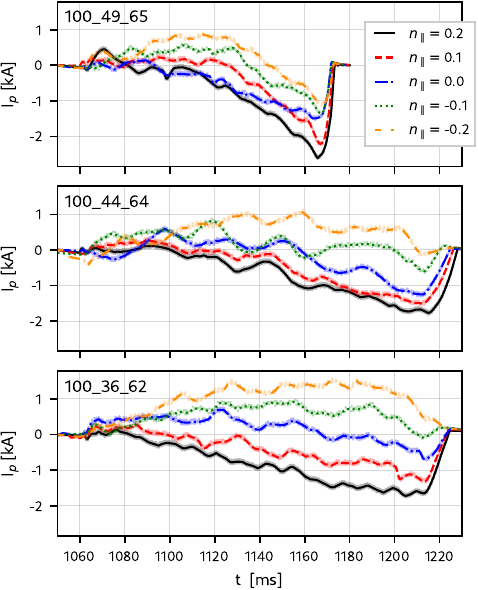}
	\caption{Plasma currents measured in the ECCD scan for the three explored magnetic configurations.}
	\label{fig:currents}
\end{figure}

Figure \ref{fig:densities} compares the time traces of the line densities when this type of scan is performed.
As expected, line densities practically overlap while plasma current variations up to 3 kA are measured at the end of the shot.
This behaviour is illustrated in figure \ref{fig:currents}.
The plasma current exhibits low frequency oscillations originated by a small ripple in the current that flows through the central and helical main coils of the device.
\begin{figure}[h] \centering
	\includegraphics[width=0.45\textwidth]{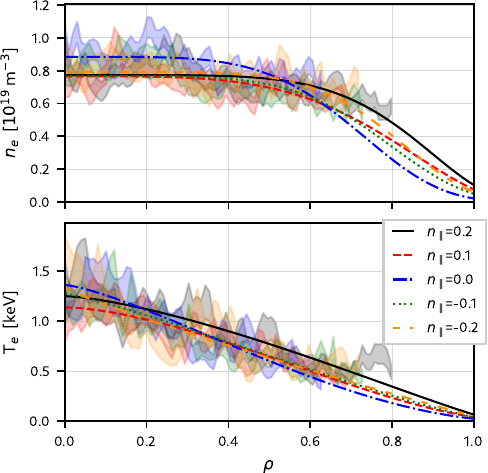}
	\caption{
		Plasma density and temperature profiles measured at $t=1185$ ms by Thomson data (colored bands) in the ECCD scan experiment for the medium \iotabar\ configuration.
		Bayesian fit to the data is also shown.
	}
	\label{fig:thomson}
\end{figure}
As we will see later in section \ref{sec:iotaevol}, following a procedure detailed in \cite{mulasValidatingNeutralbeam2023}, this inductive contribution can be extracted from the Rogowski measurement and a ``clean" current may be obtained. Counter injection with NBI2 produces negative NBCD and therefore maximum negative current is achieved when counter-ECCD is applied. Figure \ref{fig:thomson} shows the plasma density and temperature profiles in the medium iota configuration, reconstructed by integrated Bayesian analysis of Thomson scattering data (taken at \(t=1185\) ms) and profile reflectometry data \cite{vanmilligenIntegratedData2011}.
Note that they also overlap within the error bars confirming that the changes brought about by the ECCD barely impact the thermal plasma profiles.
A good density control is exhibited during the time interval of interest (between $t=1130$ and $t=1200$ ms)
for the low and medium \iotabar\ configurations.
This simplifies the mode number analysis, allows for reproducible measurements of mode potential radial structure and is fundamental when carrying out validation studies, that benefit from having a steady state fast
ion distribution and a steady state NBI2-driven current source.
In the case of the high \iotabar\ configuration, density control was harder to achieve due to degradation of wall condition, requiring an increase in ECRH1 power during NBI2 phase (up to 200 kW) and also a decrease in pulse length.

\section{Magnetic fluctuations and mode number measurements}
\label{sec:mag_fluctuations}

\begin{figure}[h] \centering
    \includegraphics[width=0.95\columnwidth]{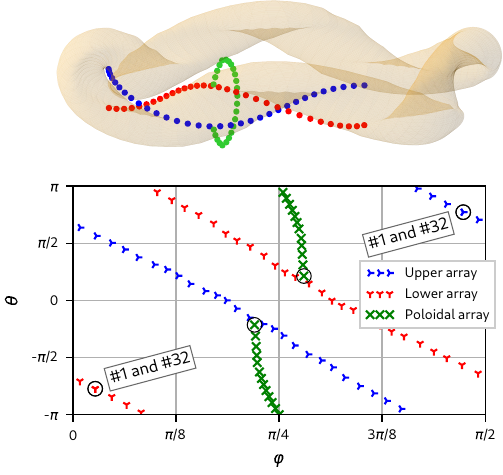}
    \caption{On top, the LCFS of TJ-II (orange) and the spatial distribution of the poloidal array coils (green) and helical array coils (red for the lower sub-array and blue for the upper one). The boozer coordinates map for each array is also shown below.}
    \label{fig:coils_setup}
\end{figure}

There are two arrays of Mirnov coils, that digitize the magnetic fluctuations at a rate of 1 MS/s.
One of them, hereinafter referred to as poloidal array, consists of 25 single-axis coils distributed in a vertical plane at constant toroidal angle, and provide information on the poloidal mode number \(m\) by measuring magnetic fluctuations in the (approximate) poloidal direction \cite{jimenez-gomezAlfvenEigenmodes2011, jimenez-gomezAnalysisMagnetohydrodynamic2007}.
The other array, hereinafter referred to as helical array, is distributed toroidally along a full period of the device, and is comprised of two up-down symmetrical sub-arrays of 32 tri-axial sensors each \cite{ascasibarMeasurementsMagnetic2022} that follow the helical path of the plasma.
These arrays are used to identify both toroidal (\(n\)) and poloidal (\(m\)) numbers of the excited modes.
Additionally, there are also two vertical arrays located at different toroidal angles not used for mode periodicity measurements.
The mode analysis is carried out using the Lomb periodogram \cite{lombLeastsquaresFrequency1976, zegenhagenAnalysisAlfven2006}, that can handle non-equispaced data points.
An in-depth discussion on the mode analysis can be found in \cite{pons-villalongaExploringOperational2024}, so here it will just be outlined.
For the mode analysis, each coil has to be mapped to a pair of magnetic angles, that in our case will be the Boozer angles of the closest point to the coil in the Last Closed Flux Surface (LCFS).
These angles are found using the vacuum equilibrium of each configuration, as tests with VMEC equilibria calculated with different plasma current profiles show that their variation is negligible in comparison with the error induced by the uncertainty in the positioning of the coils.
Figure \ref{fig:coils_setup} shows the distribution of the poloidal and helical arrays around the plasma and the Boozer angles for the medium \iotabar configuration.

\begin{figure*}[t] \centering
    \includegraphics[width=1.9\columnwidth]{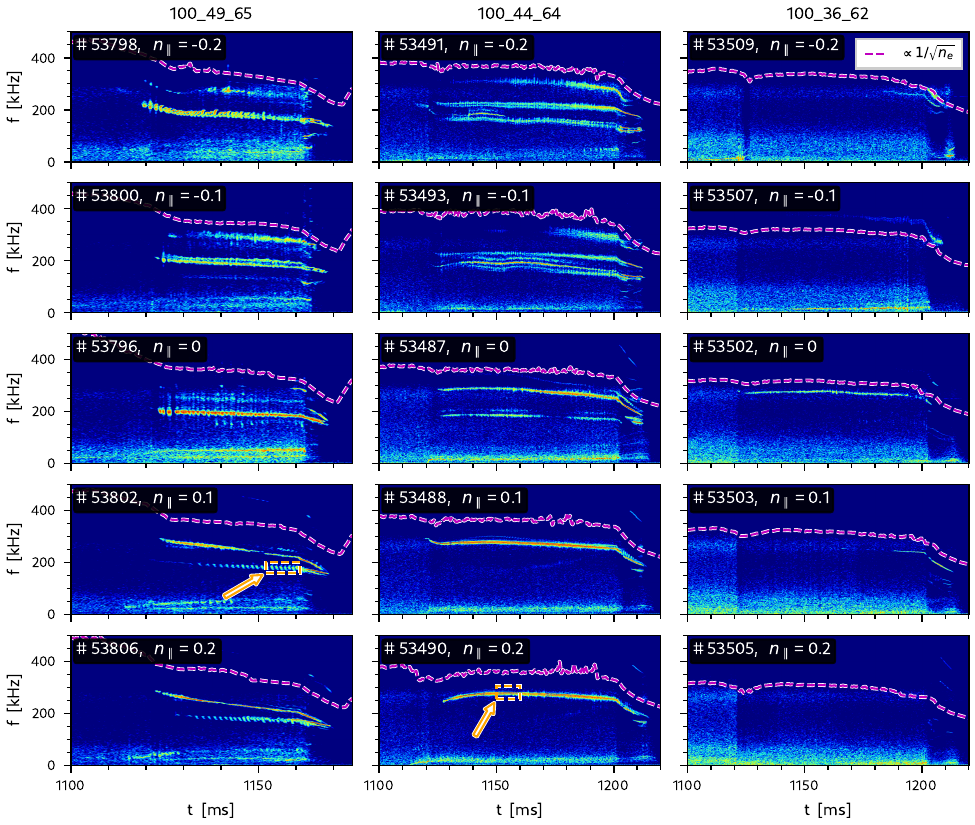}
    \caption{
        Spectrograms of magnetic fluctuations for the sweep in ECRH1 injection angle and all three magnetic configurations.
        Dashed purple lines show the (re-scaled) inverse square root of the line-integrated electron density.
        From left to right we have high \iotabar, medium \iotabar\ and low \iotabar.
        Two reproducible shots were done for each scenario. Only one of them is shown. Inside orange rectangles, marked with an arrow, the time-frequency intervals for figures \ref{fig:dmusic_compare_53802} and \ref{fig:dmusic_compare_53490}.
    }
    \label{fig:spgrams_npar_extremes}
\end{figure*}

\subsection{Observed eigenmodes}

\begin{figure}[t] \centering
    \includegraphics[width=0.4\textwidth]{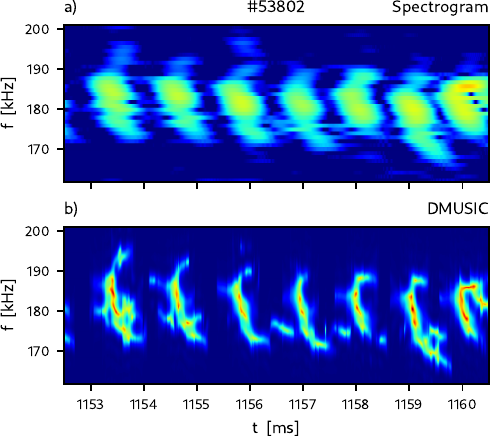}
    \caption{Comparison of STFT and DMUSIC spectrograms for shot 53802.}
    \label{fig:dmusic_compare_53802}
\end{figure}

The spectrum of NBI2 driven modes observed in each of the three magnetic configurations, for all the values of $n_{\parallel}$ explored, is shown in figure \ref{fig:spgrams_npar_extremes}.
The fact that the plasma density is not exactly constant throughout the shot allows us to identify the Alfvénic nature of the excited modes by their slow evolution in frequency, an effect which is largely amplified when the ECRH1 heating shutdown at $t=1200$ ms produces a sudden increase in plasma density and a concomitant reduction in mode frequency ($f\sim 1/\sqrt{n}$ as expected for AEs).
This figure also illustrates how the spectrum of modes is modified as the ECRH1 beam is steered along the magnetic axis.
In the medium \iotabar case, the shear Alfvén spectrum clearly evolves from a situation with three main branches well separated in frequency to a situation in which only a single intense mode is detected.
Similar changes are observed in the high \iotabar case while almost no modes are observed for the low \iotabar configuration during the ECRH1+NBI2 phase.
% Low frequency MHD modes are present in all shots, appearing less clearly in the low \iotabar\ configuration.
The set of shots illustrated in figure \ref{fig:spgrams_npar_extremes} shows a wide array of behaviours of the magnetic fluctuations.
Among the most remarkable we find:
\begin{itemize}
    \item	Modes that emerge and vanish, for example in shots 53491 and 53487, where transient modes can be observed above the \(f\sim 190\) kHz mode that is present during the entire NBI2 phase.
          This could be caused by the strong low frequency inductive oscillations in current mentioned above and the associated changes in iota or by non-linear energy transfer between the modes, both effects being outside the scope of the paper.
    \item	Transitions between chirping and stable phases, and viceversa, seen for the $n_{\parallel}=0.2$ case in shots 53806 (and 53807, not shown here). In both cases, the toroidal current reached at the end of the shot, which is the black solid line in the high \iotabar panel shown in figure \ref{fig:currents}, is maximum and negative while it also exhibits the fastest time evolution, both observations consistent with the transition from chirping to steady frequency mode \cite{melnikovTransitionChirping2016}.
    \item	Coupling of modes well separated in frequency, as happens in shot 53487, where the mode at \(\sim300\) kHz becomes more intense at \(t\simeq1175\) ms, while at the same time the mode at \(f\sim 200\) kHz weakens.
\end{itemize}
An example of chirping mode, analyzed via Short-Time Fourier Transforms (STFT) and the Damped MUltiple SIgnal Classification (DMUSIC) method is shown in figure \ref{fig:dmusic_compare_53802}.
STFT is a method that divides the signal into shorter segments, computes the Fourier transform for each segment and produces the spectrogram while
DMUSIC is a numerical tool for super-resolution frequency determination. Interested readers can find details of DMUSIC technique in \cite{kleiberModernMethods2021}.

The mode shown in figure \ref{fig:dmusic_compare_53802} displays rapid chirping behaviour, with a repetition frequency of about 1 kHz.
These chirping modes are more difficult to analyze in terms of mode numbers, as the rapid sweep in frequency deteriorates the precision of the analysis technique discussed below.

\begin{figure}[h] \centering
    \includegraphics[width=0.42\textwidth]{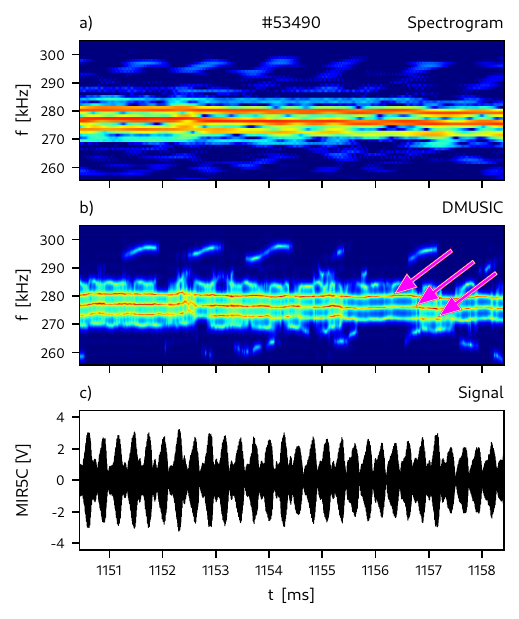}
    \caption{Comparison of STFT and DMUSIC spectrograms for shot 53490. Purple arrows highlight the three close frequency components of the mode.}
    \label{fig:dmusic_compare_53490}
\end{figure}

Aditionally, figure \ref{fig:dmusic_compare_53490} shows again a comparison of STFT and DMUSIC spectrograms calculated for shot 53490.
A closer look tells us that what looks like a single intense mode is actually formed by three distinct frequency components separated by about 4 kHz from each other, and that remain remarkably stable during the considered time interval.
The DMUSIC method is very useful in these cases, as it helps in the precise determination of the mode frequency, that is needed for the periodogram.
Figure \ref{fig:dmusic_compare_53490} also shows the bandpass filtered signal (between 250 and 310 kHz), that shows large amplitude oscillations interleaved with periods of lower amplitude as corresponds to a combination of three near-frequency modes of similar amplitude. Despite the clear frequency structure, the analysis cannot resolve different mode numbers for each of the observed frequencies. Even applying very narrow bandpass filter intended to isolate each frequency contribution the mode number result remains the same (see section \ref{sec:modeid}).

\subsection{Mode number identification}
\label{sec:modeid}

The Lomb periodogram of a set of signals \(y_{ij}\), where \(i\) is a time index and \(j\) a coil index, is given by
\begin{align}
    P(\omega, n, m) = \dfrac{1}{YY}
      & \left(
    \dfrac{\left[\sum_{ij} y_{ij} \cos(\alpha_{ij} - \tau)\right]^2}{\sum_{ij} \cos^2 (\alpha_{ij} - \tau)}
    \right. % So that the & works 
    \notag     \\
    + &
    \left.
    \dfrac{\left[\sum_{ij} y_{ij} \sin(\alpha_{ij} - \tau)\right]^2}{\sum_{ij} \sin^2 (\alpha_{ij} - \tau)}
    \right) ,
    \label{eq:LombP}
\end{align}
where \(\alpha_{ij} = m\theta + n\phi - \omega t\) is the phase term, \(\text{YY} = \sum_{ij} y_{ij}^2\) is a normalization factor, and \(\tau\) is a phase shift term, given by
\begin{equation}\label{eq:tau_definition}
    \tan 2 \tau = \dfrac{\sum_{ij} \sin 2\alpha_{ij}}{\sum_{ij} \cos 2\alpha_{ij}} \,,
\end{equation}
that makes the periodogram invariable to time and angular shifts.
The Lomb periodogram requires as input the frequency of the mode, that is obtained from the spectrogram, and the integer pair of mode numbers (\(n, m\)), noted as \(n/m\) hereinafter, over which a scan is carried out to obtain a two-dimensional distribution of intensities.
Given the direction chosen for the coil distribution arrangement and the TJ-II boozer coordinate system, the sign criterion (+) in the phase term is such that a mode number with positive \(n\) and positive \(m\) propagates toroidally along the direction of the magnetic field and poloidally along the direction of the ion diamagnetic drift.

After determining the real orientation, and correct phase sign of each set of coils by means of dedicated calibration experiments described in \cite{ascasibarMeasurementsMagnetic2022}, the measured \(\delta\dot{\mathbf{B}} (t)\) can be projected over any desired polarization vector basis (only in the case of the helical array).
In this work we will use the projection of  \(\delta\dot{\mathbf{B}} (t)\) along the normalized tangent basis vector \(\mathbf{e}_{\theta}\equiv\partial\mathbf{R}/\partial\theta\), also in Boozer coordinates, as it proved to be the best-performing basis for the mode number analysis \cite{pons-villalongaExploringOperational2024}.
A suitable collection of time and frequency intervals are selected for each mode, and the signals, calibrated and expressed in the chosen polarization basis, are then filtered and normalized to minimize distance effects on the periodogram result.
\begin{figure}[h] \centering
    \includegraphics[]{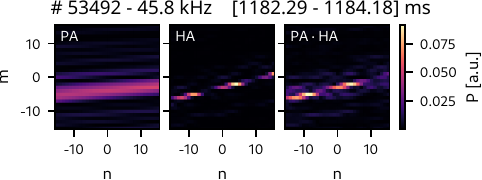}
    \caption{Example Lomb periodogram for the low-frequency mode in shot 53492. Left: poloidal array (PA), center: helical array (HA), right: product of the previous two (PA\(\cdot\)HA). The identified mode number corresponds to the maximum over this product. In this case, the analysis returns a $-9/-5$ mode.}
    \label{fig:lomb_example}
\end{figure}
\begin{figure}[h] \centering
    \includegraphics[]{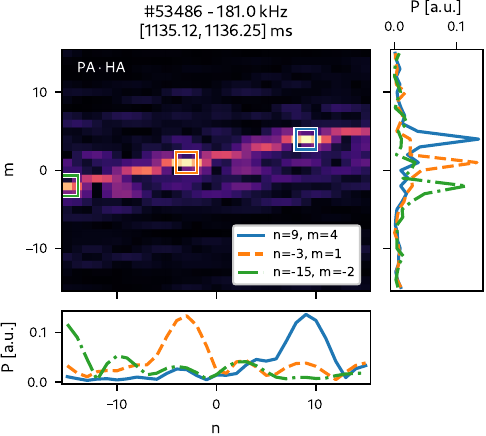}
    \caption{
        Example of periodogram (PA\(\cdot\)HA) for one of the \(n_\parallel = 0\), medium \iotabar\ shots. Several maxima can be identified (designated by squares).
        The cuts along constant \(n\) (right) and constant \(m\) (below) are also shown.
        In this time interval, both \( 9/4 \) and \( -3/1 \) appear with similar probability.
        Following the analysis procedure described in the text, that averages over several time intervals, the identified mode number is \( 9/4 \).}
    \label{fig:lomb_53486}
\end{figure}
The signals from the poloidal and helical arrays cannot be included in the same periodogram, as a difference in coil orientation induces a phase shift in the signal that compromises the accuracy of the analysis. To work around that, periodograms for the poloidal and helical arrays are computed independently and then multiplied.
Furthermore, this method also helps work around the main limitation of the helical array, that is, that the coils follow a straight path in Boozer coordinates so, in an aliasing-like phenomenon, a strip of pairs of mode numbers with slope \(1/N_\text{fp}\), that contains the true pair of mode numbers, appears in the periodogram.
The product with the periodogram of the poloidal array, that suffers to a much lesser extent from the same issue, enables in principle the determination of the true \(n/m\) pair.

\begin{figure}[h] \centering
    \includegraphics[]{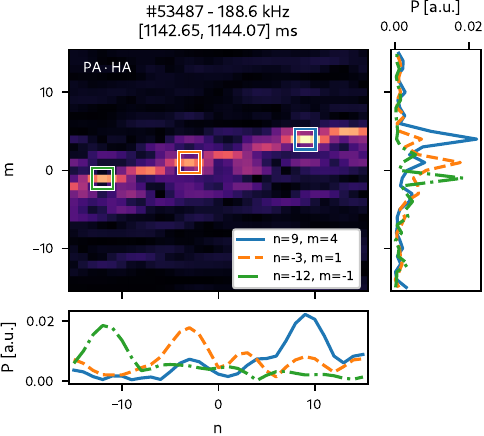}
    \caption{Another periodogram of the same mode of figure \ref{fig:lomb_53486}, this time for the other shot with medium \iotabar\ at \( n_{\parallel}=0 \).}
    \label{fig:lomb_53487}
\end{figure}

\begin{figure}[h] \centering
    \includegraphics[]{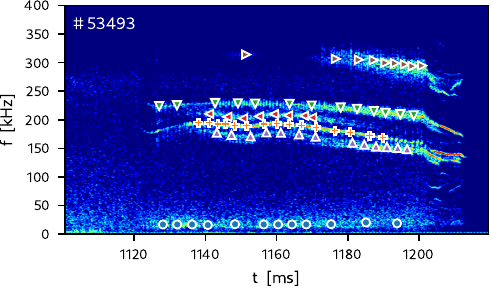}
    \caption{Time-frequency pairs used for the mode analysis of a selected shot. Different markers denote different modes. }
    \label{fig:lomb_freq_spgram_53493}
\end{figure}

\begin{figure*}[t] \centering
    \includegraphics[]{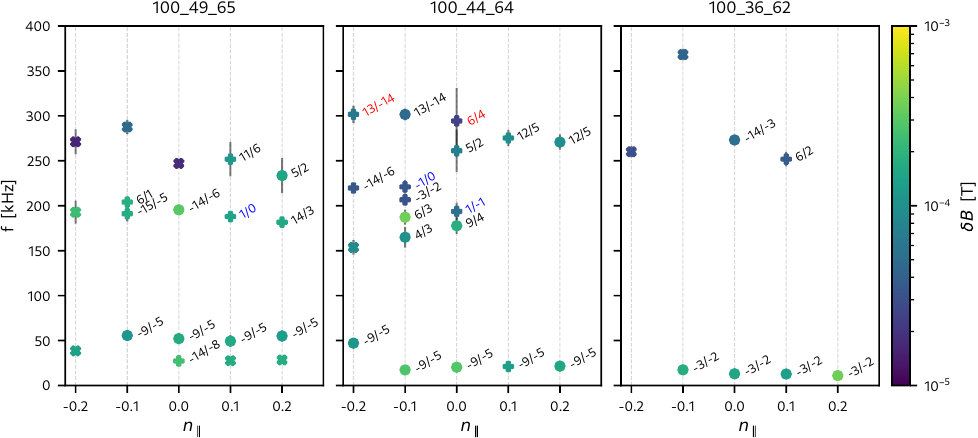}
    \caption{
        Aggregate mode analysis from all shots.
        Solid circles (\(\bullet\)) denote confidently identified mode numbers, labeled \(n/m\); plus signs (\(+\)) mark those with higher variability and lower confidence, also labeled; and product signs (\(\times\)) mark modes that could not be identified, unlabeled.
        Color shows the amplitude of the magnetic fluctuations.
        Red text denotes cases where the second most frequent dominant mode was chosen.
        Blue text denotes modes that are probably too close to the plasma core to be identified}
    \label{fig:lomb_results_aggregate}
\end{figure*}

Figure \ref{fig:lomb_example} shows an example of this process applied to a low frequency mode observed in one of the medium \iotabar\ configuration shots, with the periodograms for the poloidal array (left), helical array (center) and their product (right).
The identified pair of mode numbers is \(-9/-5\).
As we have already mentioned, a fundamental drawback due to the non-axisymmetric configuration and the spatial distribution of the coils is that there are still modes with different pairs of characteristic mode numbers that appear with almost the same probability in the analysis and we must use arguments based on the expected values of the rotational transform to discard some combinations of modes versus others.
Low frequency MHD modes have shown to be easier to identify than AEs with higher frequency even though they exhibit larger bandwidth, as is the case of the modes analyzed in this work.
The quality of the mode analysis deteriorates with rapid changes in frequency and mode intensity.
Furthermore, systematic errors during data acquisition and reduced coil availability cause a broadening of the mode number strips and sometimes introduce false positives.
High frequency broadband modes are also difficult to analyze using this method, and require more sophisticated techniques such as the Stochastic System Identification (SSI) \cite{kleiberModernMethods2021}. Especially deleterious for the analysis are spurious phase shifts \cite{horvathReducingSystematic2015}, that are also difficult to characterize experimentally for the coil arrays of TJ-II.

Typically, the difference in amplitude between the most intense \(n/m\) pair and the next is small, as can be seen in figures \ref{fig:lomb_53486} and \ref{fig:lomb_53487}, making a single analysis per mode insufficient for a confident mode identification. To circumvent this issue, many periodograms have been conducted per mode (see figure \ref{fig:lomb_freq_spgram_53493} for an example), selecting suitable time-frequency intervals over a spectrogram. This process has been conducted manually in this work but can be automatized using mode tracking algorithms \cite{vazmendesBroadbandAlfvenic2023, heinrichCharacteristicsAlfvenic2024,bustosAutomaticIdentification2021}. Then, for each mode, the most intense \(n/m\) pairs in each PA\(\cdot\)HA periodogram are selected, and the dominant mode number is identified as the most frequently occurring pair.
With this approach, we aim to reduce the uncertainties in the mode identification and improve the reliability of the results, while at the same time proceeding systematically. Figure \ref{fig:lomb_results_aggregate} shows the aggregate results of the analysis, for each configuration and \(n_\parallel\).
In some cases, the variability in the most intense mode number is too large to identify the dominant mode number, so we apply a cutoff based on the ratio between the most frequent mode number, \(N_\text{mf}\), and the total number of time intervals selected for the analysis, \(N_\text{tot}\), which is a measure of the relative frequency of occurrence of the most frequent mode number.
The following criteria are applied:

\begin{itemize}
    \item	\(N_\text{mf} / N_\text{tot} > 0.5 \ \Longrightarrow \ \) High confidence
    \item	\(0.25 < N_\text{mf} / N_\text{tot} < 0.5 \ \Longrightarrow \ \) Low confidence, but worthy of consideration after manual review
    \item	\(N_\text{mf} / N_\text{tot} < 0.25 \ \Longrightarrow \ \) Mode could not be identified
\end{itemize}

In other cases, the most frequent mode number pair lacks clear physical significance, so in those cases the second most common \(n/m\) pair is selected, provided it appears as the most intense mode a comparable number of times.
That this can happen should not be surprising, especially after inspecting figures \ref{fig:lomb_53486} and \ref{fig:lomb_53487}.
These instances are marked in figure \ref{fig:lomb_results_aggregate} by coloring the mode numbers in red.
Finally, some modes are located too close to the core to be identified \cite{pons-villalongaExploringOperational2024}.
In these cases, in which HIBP measurements can be used to distinguish true low-\(n\), low-\(m\) modes from core-localized ones (more in the next section), the mode numbers are colored blue.
\ref{app:mode_tables} contains tables with the most frequent dominant mode numbers for each scenario.

\FloatBarrier

\section{HIBP Measurements}
\label{sec:hibp}
\begin{figure}[h] \centering
    \includegraphics[width=0.8\columnwidth]{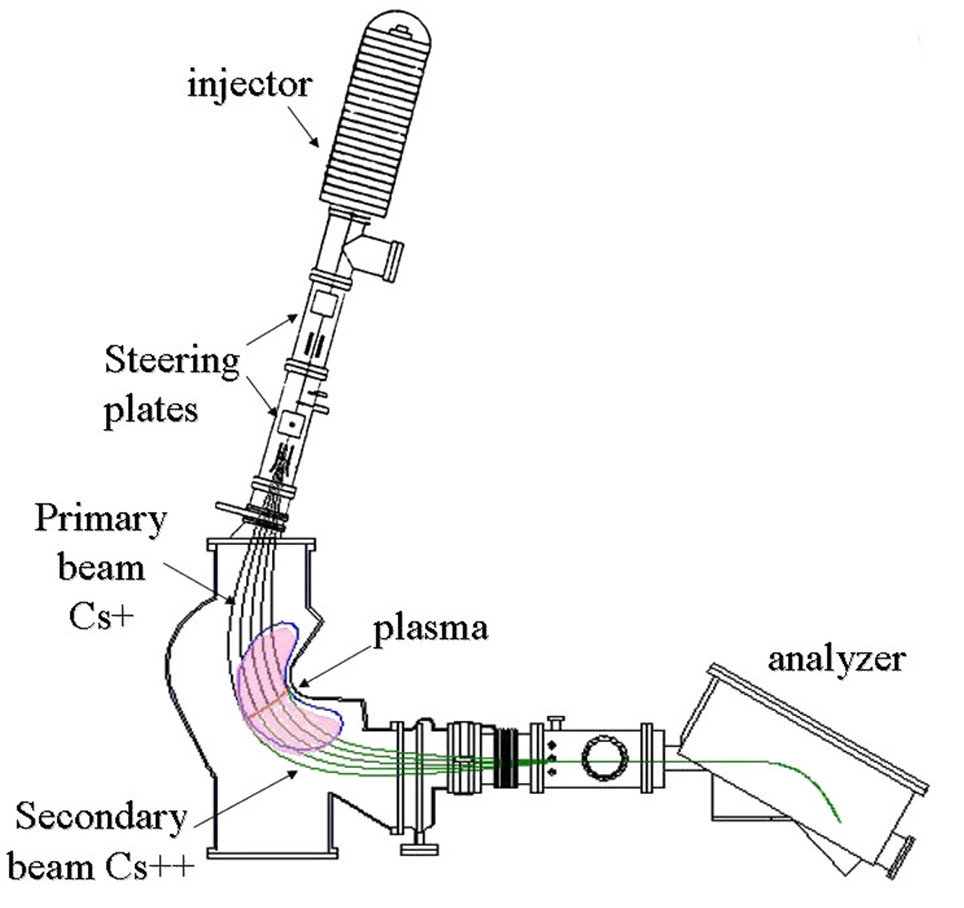}
    \caption{Schematic of the TJ-II HIBP1 diagnostic (taken from \cite{melnikovHeavyIon2017})}
    \label{fig:HIBP}
\end{figure}

Two HIBP systems (see figure \ref{fig:HIBP}) are installed in TJ-II \cite{bondarenkoInstallationAdvanced2001, melnikovHeavyIon2017}, each beam probe located in a different toroidal position. They provide data on the plasma potential, plasma density and poloidal magnetic field in the beam sample volume, that is usually swept along the plasma radius (\(t_\text{sweep} \sim 10-15\) ms) so as to provide profiles of these quantities.
In the absence of magnetics diagnostics, or when this latter are no longer reliable (core modes), the ability to measure the local poloidal field at different spatially close sample volumes can also be exploited to provide a rough estimate of the poloidal mode number \cite{melnikovInternalAEs2010}.

\begin{figure}[h] \centering
    \includegraphics[width=0.9\columnwidth]{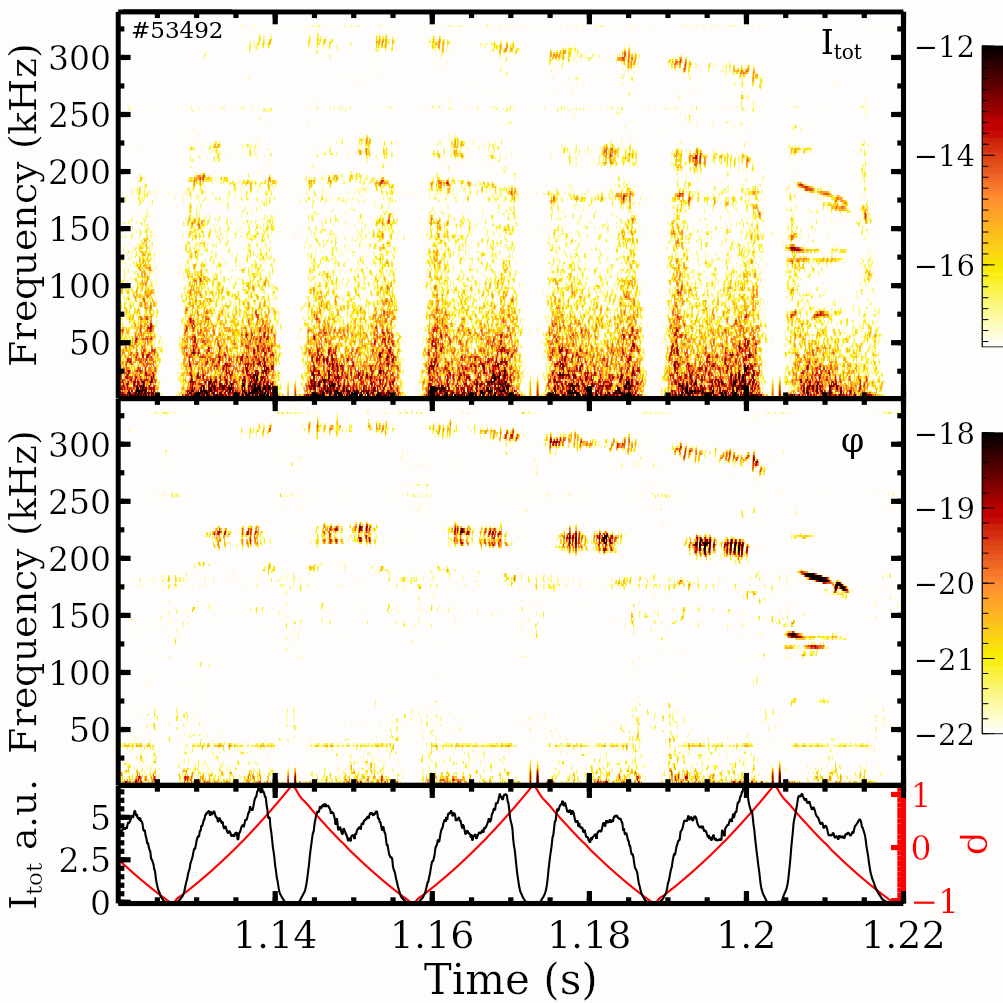}
    \caption{Spectrogram of the heavy ion beam intensity $I_{tot}$ and plasma potential $\Phi$.
        The evolution of $I_{tot}$ as the beam scans the measurement positions is shown in the bottom panel.
        The corresponding $\rho$ positions are also shown (red line).
        The data corresponds to one of the $n_{\parallel}=-0.2$ shots (\#53492, medium \iotabar).}
    \label{fig:HIBP_example_mes}
\end{figure}

Of particular interest for this work are the plasma potential perturbation profiles, that can be used to estimate the radial positions of the measured AEs.
In this case HIBP1 was used in radial scan mode while the other probe (HIBP2) was only operated in some of the shots in fixed point modality, measuring plasma potential at $\rho=0.6$.
Figure \ref{fig:HIBP_example_mes} shows an example of the measured quantities for the $n_{\parallel}=-0.2$ case in the medium \iotabar\ configuration (shot \#53492).
The spectrogram of the magnetic fluctuations corresponding to a shot with the same conditions (\#53491) is shown in figure \ref{fig:spgrams_npar_extremes}.
The more or less stationary plasma, which translates into a constant mode frequency, allows us to average over repetitive scans of the heavy ion beam and retrieve a radial profile of the mode.
\begin{figure}[h] \centering
    \includegraphics[width=0.7\columnwidth]{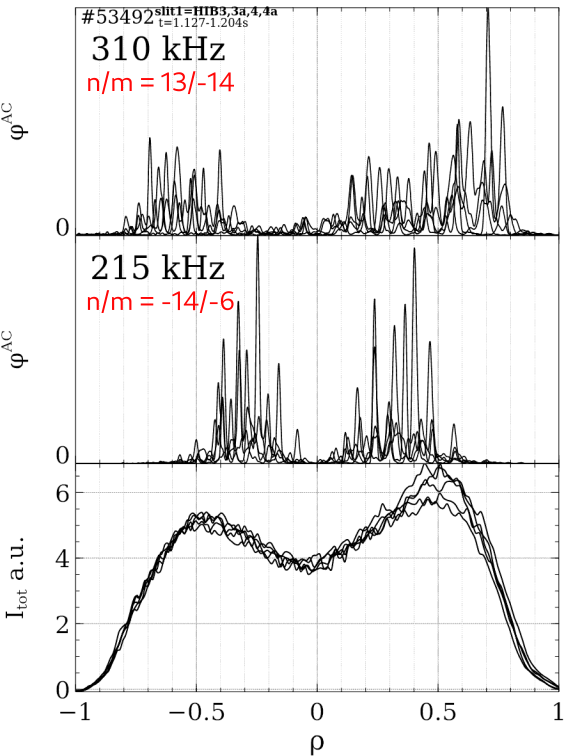}
    \caption{Radial profile of mode potential perturbations for $n_{\parallel}=-0.2$ and medium \iotabar\ (\#53492 and \#53491). The results of six complete radial scans of the heavy ion beam are overplotted.}
    \label{fig:HIBP_mode_profiles_53492}
\end{figure}
This result, that can already be guessed by inspecting the spectrogram of plasma potential fluctuations presented in figure \ref{fig:HIBP_example_mes} is shown in figure \ref{fig:HIBP_mode_profiles_53492}.
In this case, the spectrogram of magnetic fluctuations shows three distinct modes around 180, 215 and 310 kHz.
The one with lower frequency is barely seen by HIBP1 in either density or potential fluctuations.
The radial profile measurements reveal a more core localized 215 kHz mode with an approximately symmetrical profile while the 310 kHz mode is shifted towards the medium region of the plasma and exhibits of somewhat asymmetrical profile.
The corresponding mode numbers extracted from the Mirnov coils data were given in figure \ref{fig:lomb_results_aggregate} and appear also in the figures.
The bottom panel in figure \ref{fig:HIBP_mode_profiles_53492} shows the dependence of the secondary beam intensity (Cs$^{++}$) on the radial position.
As shown in reference \cite{khabanovDensityProfile2019}, the density profile, roughly proportional to the beam intensity profile can be reconstructed from this measurement.
\begin{figure}[t] \centering
    \includegraphics[width=0.7\columnwidth]{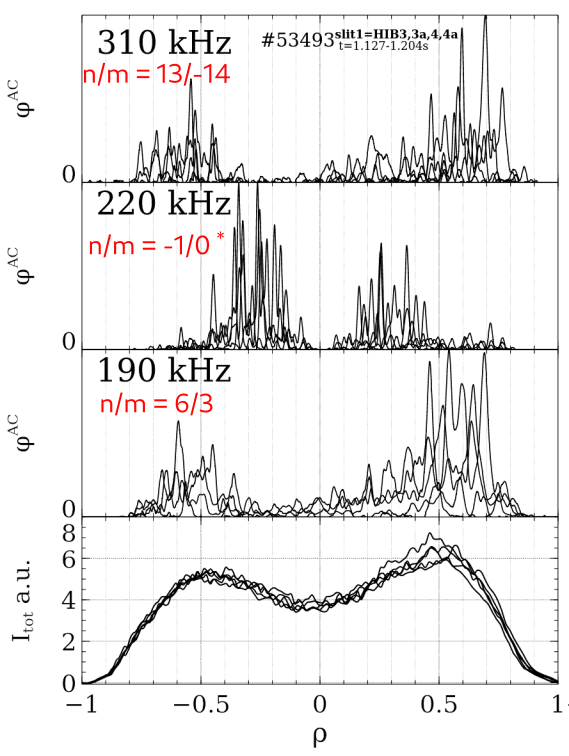}
    \includegraphics[width=0.7\columnwidth]{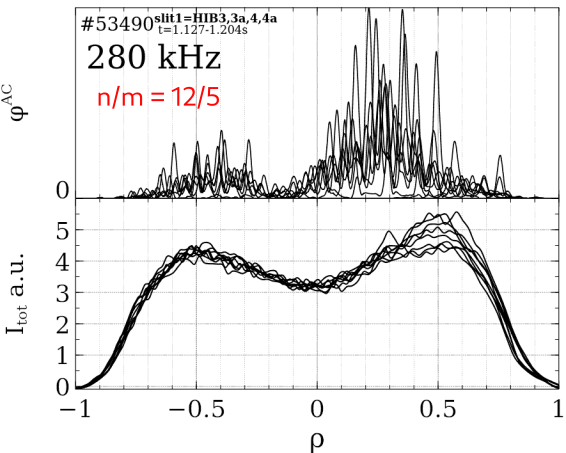}
    \caption{Radial profile of mode potential perturbations measured at medium \iotabar\ for $n_{\parallel}=-0.1$ (\#53493) and $+0.2$ (\#53490).
        The profile of the $n_{\parallel}=+0.2$ case corresponds to the intense structure of three close frequency modes that is shown in figure \ref{fig:dmusic_compare_53490}.
        The \(n/m = -1/0\) mode at 220 kHz is located too close to the plasma core, and mode numbers cannot be correctly identified using the Mirnov arrays.
    }
    \label{fig:HIBP_mode_profiles_53493_90}
\end{figure}
Note that, consistently with the density profiles shown in figure \ref{fig:thomson}, the intensity gradient region falls between $\rho=0.5$ and $\rho=0.8$.
This is important when considering the excitation of low frequency MHD modes and its relation to the rotational transform profile that will be explored in section \ref{sec:iotaevol}.

\begin{figure}[t] \centering
    \includegraphics[width=0.9\columnwidth]{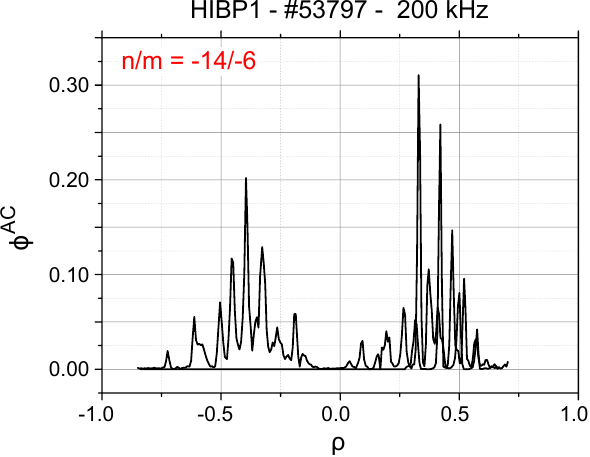}
    \caption{HIBP potential profile for the high iota configuration and $n_{\parallel}=0$ (\#53797).}
    \label{fig:HIBP_48__53797__200khz}
\end{figure}

As mentioned above, for some modes a radial asymmetry in the potential profile is clearly observed.
This is consistent with the fact that these are probably modes that are destabilized thanks to the existence of gaps in the continuum of the shear Alfvén spectrum \cite{kolesnichenkoAlfvenEigenmodes2002}.
These modes consist of coupled modes with different \(m\)'s and \(n\)'s that evolve together. The difference in poloidal mode numbers produces the asymmetry in the potential profiles.
\begin{figure}[h] \centering
    \includegraphics[width=0.9\columnwidth]{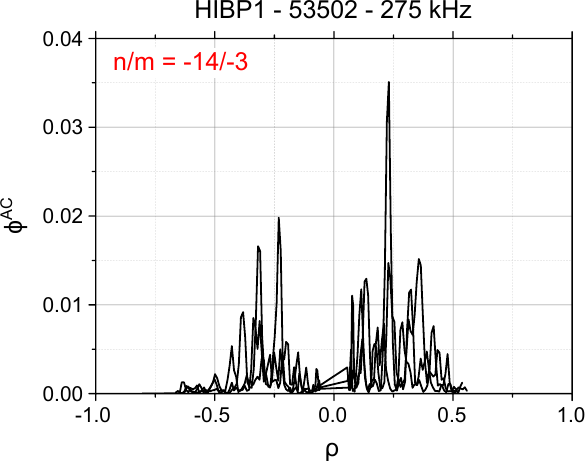}
    \caption{HIBP potential profile for the low iota configuration and $n_{\parallel}=0$ (\#53502).}
    \label{fig:HIBP_36__53502__275khz}
\end{figure}
On the other hand, as it was demonstrated in reference \cite{pons-villalongaExploringOperational2024} using a synthetic Mirnov coils diagnostic, in the case of coupled modes, the magnetic coils cannot resolve different $m$'s when the difference between them is $\Delta m=\pm 1$ or $\pm 2$.
In these case, only the dominant mode numbers result from the analysis.
The radial profiles of the mode potential have been measured for several values of $n_{\parallel}$ and additional results for $n_\parallel=-0.1$ and $+0.2$ in the medium \iotabar\ onfiguration are shown in figure \ref{fig:HIBP_mode_profiles_53493_90}.
The results for perpendicular injection ($n_{\parallel}=0$) in the low and high \iotabar\ onfigurations are shown in figures \ref{fig:HIBP_36__53502__275khz} and \ref{fig:HIBP_48__53797__200khz} respectively.
The shorter pulse length achievable in the high \iotabar\ case and the low amplitude of the mode observed in the low \iotabar\ configuration produce less robust measurements of the perturbation radial profile.

\section{Time evolution of rotational transform}
\label{sec:iotaevol}
To understand the changes that occur in the Alfvén wave spectrum when we modify the direction of the microwaves beam it is necessary to have an estimate of the time evolution of the rotational transform.
When this evolution is due to a current source that is established from almost the beginning or when we observe that the value of the toroidal current measured by the Rogowski coil stabilizes before the end of the shot we can assume that, in the final part of the shot, the plasma shielding current is zero and the current profile inside the plasma is that of the current source that we force in the plasma.
These sources are generated by ECCD, NBCD or by the bootstrap current inherent to the configuration and the plasma profiles.
With this approximation we can have an estimate of the rotational transform at the end of the shot \cite{cappaStabilityAnalysis2021}.
However, we cannot use this approach safely in the experiments discussed in this paper because we have a source of ECCD current during the whole shot duration, and a second source due to NBI2 which is established several milliseconds after the neutral beam injection.
Furthermore we observe in most of the cases that the current is not stabilized at the end of the shot.

The time evolution of the toroidal current $I(\rho,t)$, and consequently of \iotabar , can be estimated using a simple model to describe the radial diffusion of the sum of the different current sources $I_s(\rho)$ in a cylindrical plasma.
To this end we need to solve the evolution equation for the shielding current $I_E(\rho,\tau)$, which is derived in \ref{app:rad_diffusion} (equation \ref{eq:adim_radialdif}). Once $I_E(\rho,\tau)$ is known, the rotational transform profile can be calculated using equation \ref{eq:iota_evol}.

The toroidal current source $I_s(\rho)$ is a combination of the current induced by ECCD, NBCD and bootstrap current.
In these low density plasmas the bootstrap current is smaller than the currents due to the heating systems, it is not centered at the core, its impact on \iotabar\ is therefore smaller, and we will not take it into account in the calculation.
Also, we assume that the ECCD and NBCD current sources are not time-dependent.
This assumption is reasonable since we have an approximately constant density throughout the shot.
In addition, we also assume a constant conductivity over time, but which may not be the same depending on whether we are in the ECRH or ECRH+NBI phase due to changes in plasma temperature.
In fact, as we see shortly, the complete time evolution must be calculated in two phases whenever an ECCD contribution ($n_{\parallel}\neq 0)$ is present.

\subsection{Initial conditions}

According to expressions \ref{eq:current_dens} and \ref{eq:int_current_dens}, the shielding current $I_E$ is related to the total plasma current $I$ as follows
\begin{equation}
    I_E(\rho,\tau)=I(\rho,\tau)-I_s(\rho)
\end{equation}
and therefore the initial condition for $I_E$ is given by
\begin{equation}
    I_E(\rho,0)=  I(\rho,0)-I_s(\rho) \label{eq:init}
\end{equation}
We stated above that the main current sources come from ECCD and NBCD, that is $I_s(\rho)\equiv I_{\text{ECCD}}(\rho)+I_{\text{NBCD}}(\rho)$.
For both the ECCD current at different injection angles and the NBCD current produced by the counter injector we will use a combination of estimates presented in previous works together with the plasma toroidal current measured in the experiments discussed here.

\subsubsection{NBCD current source.}

\begin{figure}[h]
    \centering
    \includegraphics[width=\linewidth]{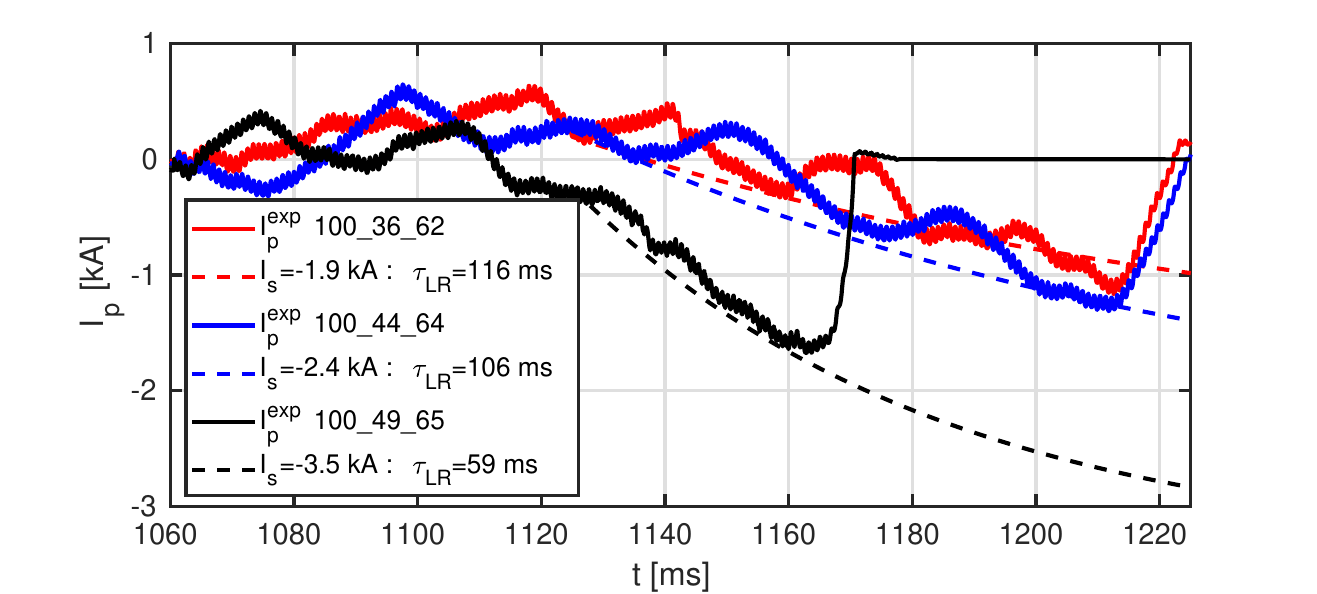}
    \caption{Plasma current measured in low (solid red line), medium (solid blue line) and high (solid black line) \iotabar\ configuration for $n_{\parallel}=0$. The current evolution once the inductive oscillations have been removed is represented by the dashed lines.}
    \label{fig:current_sources_npar0}
\end{figure}

The NBCD profile has been taken from the \texttt{ASCOT5} \cite{hirvijokiASCOT2014} slowing-down simulations presented in \cite{mulasASCOT5Simulations2022}.
The total NBCD current obtained in the medium \iotabar\ configuration was validated against experimental results in \cite{mulasValidatingNeutralbeam2023} and a good agreement was observed for neutral beam counter-injection at line densities and density profiles similar to the ones shown in figures \ref{fig:densities} and \ref{fig:thomson}.
For our present purposes, and having in mind that different configurations are used and that the plasma densities slightly differ between the three of them, the integrated current of the NBCD source will be determined from the plasma toroidal current measured in the $n_{\parallel}=0$ cases and we will take the current density profile calculated in reference \cite{mulasValidatingNeutralbeam2023}.
We mentioned in section \ref{sec:exp_setup} that the low frequency oscillations originated by inductive coupling with the main conductors can be extracted from the Rogowski coil measurement.
Note that these oscillations not only occur with respect to a certain average value, but may also be shifted upwards or downwards depending on the specific behavior of current ripple in the central and helical coils of the device.
Applying the procedure described in the appendix of \cite{mulasValidatingNeutralbeam2023} to the current measured in the NBI2 phase we arrive at the result shown in figure \ref{fig:current_sources_npar0}.
The values of the integrated NBCD current source (no ECCD in this case) in the limit of zero shielding current and the LR time for each case are given in the figure.
We note here for later, when we will discuss the boundary conditions given by equation \ref{eq:bound}, that the values of $\tau_{LR}$ used to solve equation \ref{eq:adim_radialdif} are determined experimentally from the exponential time evolution of the current once the oscillations have been removed.
The current measured in the medium and high iota configurations (actually the limit current inferred from the measurements) agrees well with the results presented in \cite{mulasValidatingNeutralbeam2023}.
However, the NBCD measured in the low iota configuration appears to be lower.
A possible reason for this is the slightly lower plasma volume and size of this configuration (see figure  \ref{fig:iota}) and also the fact that plasma line densities in this case ($\sim0.5\times 10^{19} \text{ m}^{-3}$) were a bit lower than the line densities achieved in the other two ($\sim 0.6-0.7\times 10^{19}\text{ m}^{-3}$).
The beam shine-trough power and subsequent slowing-down \texttt{ASCOT5} simulations that would clarify these differences are out of the scope of this work.

\subsubsection{ECCD current source.}

\begin{figure}[h]
    \centering
    \includegraphics[width=\linewidth]{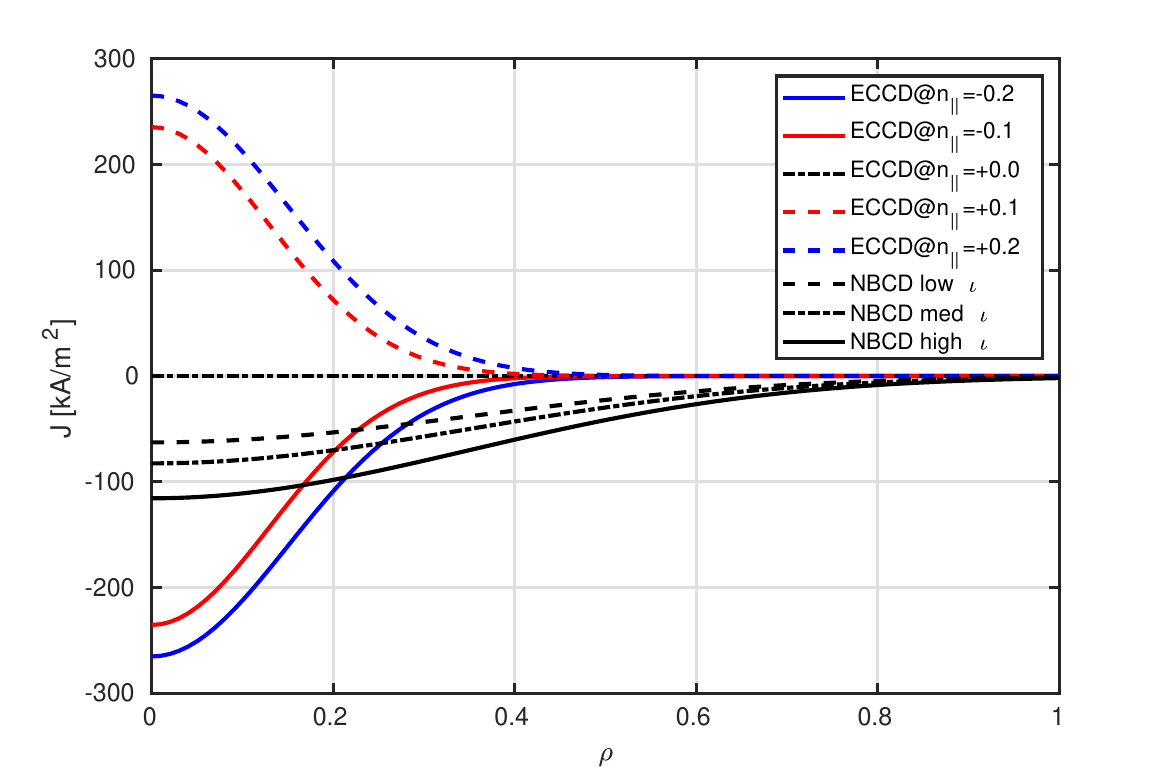}
    \caption{ECCD and NBCD current densities used in the \iotabar\ evolution simulations. NBCD source is configuration dependent in a manner consistent with the experimental observations and ECCD magnitude and profile width depends on the value of $n_{\parallel}$.}
    \label{fig:current_sources}
\end{figure}

The ECCD profile is estimated from experimental observations of ECCD total induced current in oblique launching experiments \cite{fernandezECCDExperiments2008} and fast power modulation experiments \cite{eguiliorHeatWave2003} for perpendicular injection combined with the results of ray-tracing and beam tracing simulations \cite{martinez-fernandezHighPower2020}.
We use these simulations to determine how the power deposition and ECCD profiles widens as $|n_{\parallel}|$ increases.
This result is corrected with the measured power deposition profile at $n_{\parallel}=0$ by imposing that its integral is the total measured current after several LR times.
With these assumptions we obtain the current densities that are shown in figure \ref{fig:current_sources}.

\subsection{Boundary conditions}

The boundary condition for $I_E$ is given by
\begin{equation}
    I_E(1,\tau)=  I(1,\tau)-I_s(1).   \label{eq:bound}
\end{equation}
We see that the quantity $I(1,\tau)$ is nothing else than the experimental toroidal current $I_{p}(\tau)$ measured by the Rogowski coil that is shown in figure \ref{fig:currents}.
For stationary current sources, the evolution of the toroidal current can be expressed as
\begin{equation}
    I_p(\tau)=I_s(1)(1-e^{-\tau})+I_p(0)e^{-\tau}
    \label{eq:Itor_exp}
\end{equation}
and the boundary condition takes the final form
\begin{equation}
    I_E(1,\tau)=[I_p(0)-I_s(1)]e^{-\tau}
\end{equation}
Since all of the above is only valid for stationary current sources, the complete time evolution when $n_{\parallel}\neq 0$ must be calculated in two phases; first with an ECCD current source starting from zero plasma current and then with a combined ECCD+NBCD current source starting from the final plasma current profile given by the first simulation.
Note that since $\tau=t/\tau_{LR}$, the time constant of the plasma column ($\tau_{LR}$) is actually determined experimentally by fitting the measured time evolution of the plasma current to its theoretical evolution given
by expression \ref{eq:Itor_exp}.
In this case, following the definition of $\tau_{LR}$ of a cylindrical infinite plasma column given in \ref{app:rad_diffusion} ($\tau_{LR}\equiv\mu_0 a^2/\eta_0$), this is equivalent to determine a central resistivity for the cylindrical plasma such that its $\tau_{LR}$ fits the experimental evolution of the toroidal current.
We do not intend to compare this result with the actual plasma conductivity that may obtained from neoclassical theory since neither the inductances of both infinite cylindrical and real plasma columns, nor the expressions for $\tau_{LR}$, are the same.

\subsection{Iota evolution results without ECCD \((n_{\parallel}=0)\)}

\begin{figure}[t]
    \centering
    \includegraphics[width=0.85\linewidth]{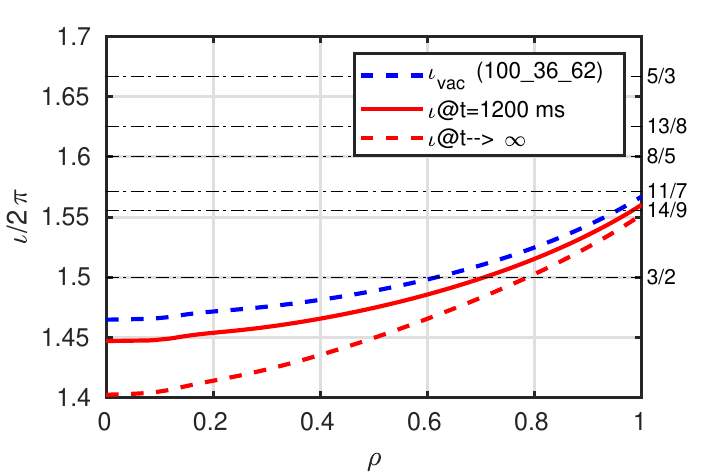}\\
    \includegraphics[width=0.85\linewidth]{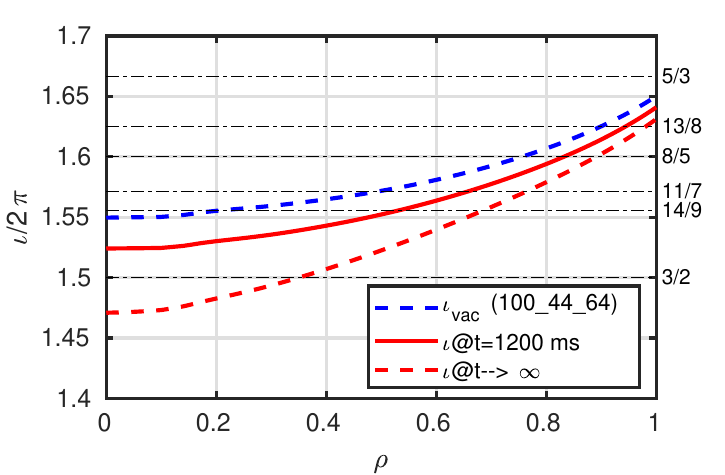}\\
    \includegraphics[width=0.85\linewidth]{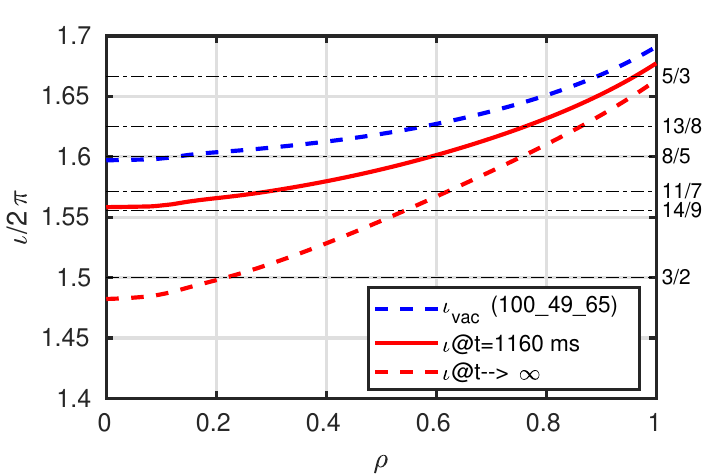}\\
    \caption{Rotational transform profiles at $t=1200$ ms and $t=1160$ ms (solid red lines) for the low, medium and high \iotabar\ respectively. The vacuum (blue dashed lines) \iotabar\ and the equilibrium ($t \rightarrow\infty$) \iotabar\ (dashed red lines) are shown for each magnetic configuration.}
    \label{fig:nbcd_npar0}
\end{figure}

The result of applying this model to the cases with no ECCD current (\(n_{\parallel}=0\)) are shown in figure \ref{fig:nbcd_npar0}.
The calculated evolution shows that for these cases in which the toroidal current is still evolving (see figure \ref{fig:current_sources_npar0}), the profile of \iotabar\ at the end of the shot (\(t = 1200\) ms) is midway between the vacuum value and the equilibrium value.
It is enlightening to compare the \iotabar\ profiles shown in the figures with the mode numbers of the low frequency MHD instabilites (\(f<100\) kHz) shown in figure \ref{fig:lomb_results_aggregate}.
Restricting ourselves to the shots with \(n_{\parallel}=0\) for the moment, the mode number analysis shows a \(-3/-2\) instability for the low \iotabar\ configuration, a \(-9/-5\) for the medium \iotabar\ case and two instabilites with mode numbers \(-9/-5\) and \(-14/-8\) for the high \iotabar\ case.
This is consistent with the calculated \iotabar\ profiles considering that this type of instability is triggered in the density gradient region ($\rho \approx 0.5-0.8$) in the presence of a low order rational value; $\rho_{3/2}\approx 0.7$, $\rho_{8/5}\approx 0.8$ and $\rho_{8/5}\approx 0.6$, $\rho_{13/8}\approx 0.75$ for the low, medium and high \iotabar\ respectively.
Note that following the numerical study of Mirnov coils performance presented in \cite{pons-villalongaExploringOperational2024} the calculation of the mode numbers is subject to an indeterminacy of $\pm 1$ or greater depending on the radial position of the instability and thus the pairs $-9/-5$ and $-14/-8$ delivered by the analysis may very well correspond to $-8/-5$ and $-13/-8$ having in mind that actually $\iotabar=9/5$ and $\iotabar=14/8$ are well outside the reasonable \iotabar\ values when negative NBCD current is at play.

\subsection{Iota evolution results with ECCD}

\begin{figure}[t]
    \centering
    \includegraphics[width=0.85\linewidth]{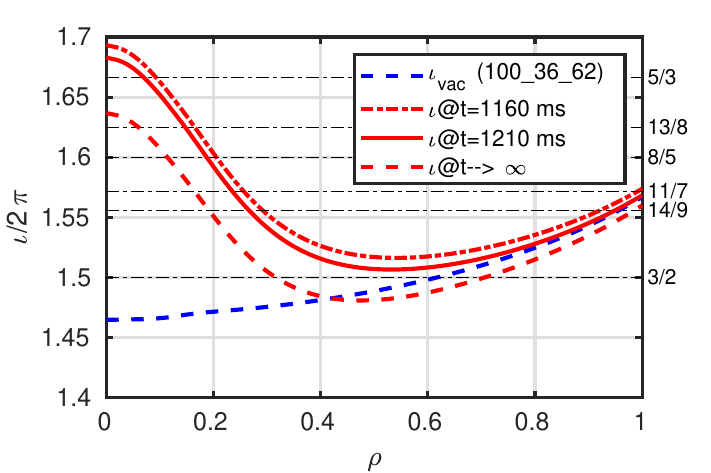}\\
    \includegraphics[width=0.85\linewidth]{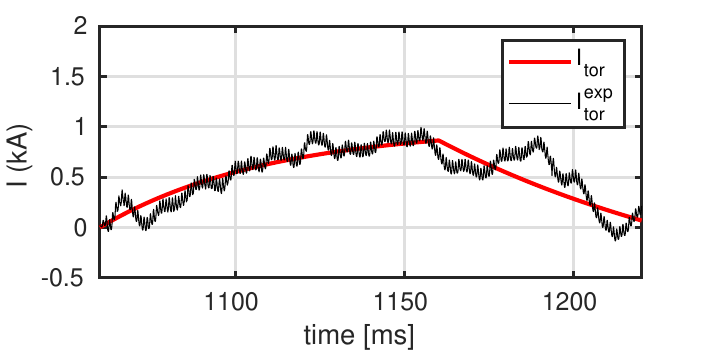}
    \caption{Low \iotabar\ configuration with $n_{\parallel}=-0.1$. On top, we show the rotational transform profiles at the beggining of the ECRH1+NBI2 phase ($t=1160$ ms), when ECCD is dominant (dashed-dot red lines) and at the end of the shot ($t=1210$ ms) when NBCD has lowered the central part of the \iotabar\ profile so that the 3/2 almost appears in the plasma (solid red lines). The vacuum (blue dashed lines) \iotabar\ and the equilibrium ($t \rightarrow\infty$) \iotabar\ (dashed red lines) are shown. Bottom panel shows the time evolution of the experimental (black line) and modeled (red line) total currents.}
    \label{fig:nbcd_npar-0.1}
\end{figure}

Let us now look at how this low frequency activity changes as a function of the ECCD current amplitude.
Take for instance the $-3/-2$ mode observed in the low \iotabar\ case.
For even more negative current, that is, adding counter-ECCD with $n_{\parallel}>0$ to the already negative NBCD, the $-3/-2$ mode is always present with more or less amplitude (that probably depends of the position of the rational in respect to the density gradient).
For maximum positive ECCD (\(n_\parallel=-0.2\)), \(\iotabar=3/2\) disappears since positive current elevates the \iotabar\ profile.
The most interesting case, for which the calculation of the time evolution of iota may give us a clue, is the one for $n_{\parallel}=-0.1$ shown in figure \ref{fig:nbcd_npar-0.1}.
In this case, co-ECCD pushes-up the \iotabar\ profile but at soon as NBCD comes into play, its contribution makes the $3/2$ rational surface appear again.
\begin{figure}[h]
    \centering
    \includegraphics[width=0.85\linewidth]{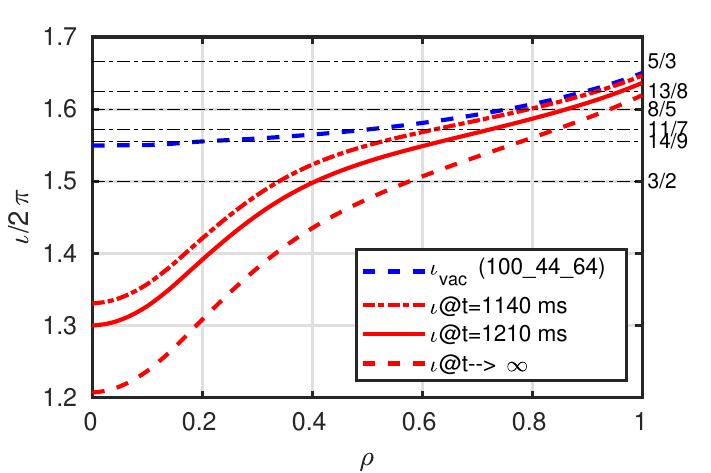}\\
    \includegraphics[width=0.85\linewidth]{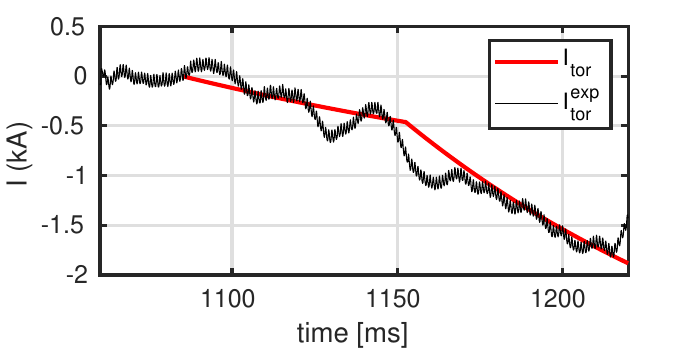}
    \caption{Same as \ref{fig:nbcd_npar-0.1} for the medium \iotabar\ configuration with $n_{\parallel}=+0.2$. ECRH1+NBI2 phase starts now at ($t=1140$ ms).}
    \label{fig:nbcd_npar0.2}
\end{figure}
This is also the case for which the \(-3/-2\) perturbation shows the largest amplitude in the magnetic signal.
In figure \ref{fig:nbcd_npar-0.1}, both the final iota profile due to ECCD current only and the final profile due to the combination of both sources are presented.
The first one does not actually contain the $3/2$ low order rational but the second one ($t\rightarrow\infty$) intersects $\iotabar=3/2$ at $\rho\approx 0.7$ and $\rho\approx 0.32$.
The two-step model cannot capture exactly the time sequence of the experiment but clearly illustrates the phenomenon.
With the current parameters of this cylindrical model we cannot reproduce exactly the experimental behaviour but the main trends are well illustrated.
Note that the combination of both negative and positive current sources, one broader than the other, produce a minimum in the \iotabar\ profile.

The same reasoning can be applied to the low frequency instabilities observed for the other two magnetic configurations.
For instance, figure \ref{fig:nbcd_npar0.2} shows a case of maximum negative current ($n_ {\parallel}=0.2$) for the medium \iotabar\ configuration.
The $8/5$ rational surface is located around $\rho=0.85$, consistently with the \(-9/-5\) pair determined by the magnetic arrays.
According to the model result, even for this maximum negative current the 3/2 rational is still in the inner plasma.
The fact that we observe no low frequency mode with consistent $-3/-2$ pair, as the ones observed for the low \iotabar case, supports the validity of the model.

\subsection{MSE measurements}

At the time of carrying out these experiments, the MSE diagnostic was not available and the measurements presented here were performed a posteriori in the medium \iotabar\ configuration only, using a very similar plasma density.
%\begin{figure}[h]
%    \centering
%    \includegraphics[width=0.7\linewidth]{s05/mse_setup-eps-converted-to.pdf}
%    \caption{On top we show the magnetic surfaces and field contours corresponding to the vacuum \texttt{VMEC} equilibrium of the medium \iotabar\ configuration.
%        The part of the diagnostic beam trajectory seen by the spectrometer is represented (dashed black line). Bottom panel shows the values of $\alpha$ along the beam.}
%    \label{fig:mse_setup}
%\end{figure}
%The basic set-up and the values of $\alpha$ for the medium iota configuration in vaccum are shown in figure \ref{fig:mse_setup}.
\begin{figure}[h]
    \centering
    \includegraphics[trim={0 1cm 0 0.7cm},width=0.95\linewidth]{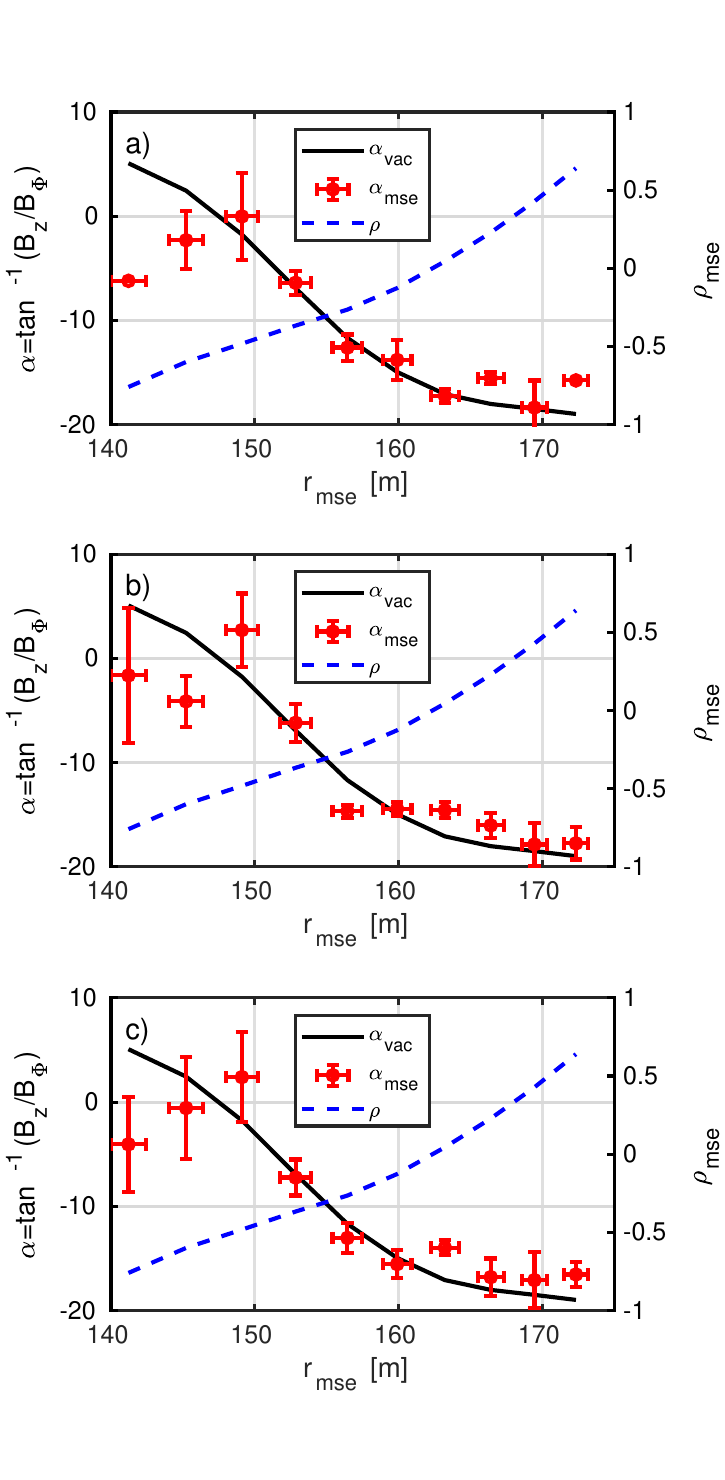}
    \caption{Angle $\alpha$ between magnetic field components measured by MSE for perpendicular ECRH injection (a) co-injection with $n_{\parallel}=-0.2$ (b) and counter-injection with $n_{\parallel}=+0.2$ (c). The vaccum theoretical values (black solid lines) and the relation between $\rho$ and $r_{mse}$ (dashed blue lines) are also shown.}
    \label{fig:mse_data}
\end{figure}
Plasma temperature was higher since, due to poor wall conditioning, the ECRH2 beam had to be maintained throughout the shot to help control the plasma density.

MSE measures the orientation of the magnetic field vector perpendicular to the injection direction of the diagnostic neutral beam \cite{mccarthySpectrallyResolved2015}.
This information is usually given in terms of the angle $\alpha(r_{mse})\equiv\arctan(B_z(r_{mse})/B_{\phi}(r_{mse}))$, where $r_{mse}$ are the positions along the beam where the lines of sight of the spectrometer intersect the beam.
Using the \texttt{VMEC} code to calculate the plasma equilibrium and its \iotabar\ profile for a given profile of toroidal current, and determining the components of the field perpendicular to the beam, we may estimate the quantity $\alpha$ measured by MSE and establish a correspondance with the \iotabar\ profile.
Figure \ref{fig:mse_data} shows the results of the MSE measurements for different launching angles of the ECRH1 beam.
These were made at the end of the NBI2+ECRH(1\&2) phase where both NBCD and ECCD are acting to modify the rotational transform.
Although MSE can clearly distinguish between rotational transforms of different configurations \cite{mccarthySpectrallyResolved2015}, it cannot resolve changes in $\alpha$ due only to different plasma currents.

\begin{figure}[h]
    \centering
    \includegraphics[width=0.8\linewidth]{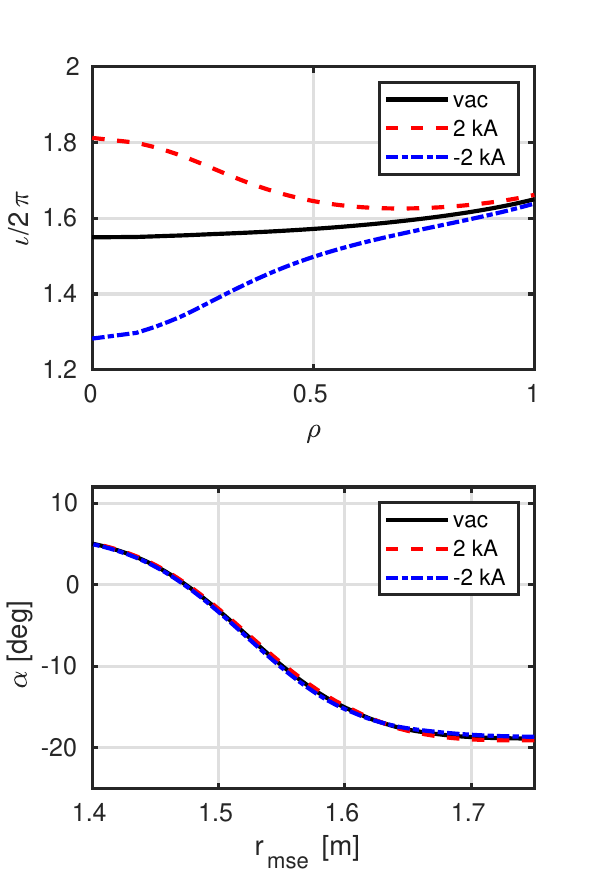}
    \caption{Synthetic MSE measurements in vaccum (black solid lines) compared to the result that would be obtained wth co- and counter-ECCD current. The impact on the angle $\alpha$ is barely noticeable for such small currents.}
    \label{fig:mse_synth}
\end{figure}

Actually, this can be confirmed calculating the expected $\alpha$ for different \texttt{VMEC} equilibriums modified by toroidal current.
Figure \ref{fig:mse_synth} compares the profiles of \iotabar\ and $\alpha$ without plasma current and for core localized ECCD current densities with $+2$ kA and $-2$ kA.
It is shown that even though the changes in current are relevant from the point of view of Alfvén modes excitation, they are not large enough to be detected by MSE, whose error bars greatly exceed the variations expected from changes in plasma current.

%\FloatBarrier

\section{Discussion}
\label{sec:discuss}

Similarly to what happens in tokamaks, where modes of the same $n$ and different $m$ couple within the TAE gap, the existence of helicity gaps (HAE) in stellarators allows for complex radially extended and weakly damped structures characterized by modes with different values of $n$ and $m$.
Gaps in the continuum, that favors the existence of radially extended discrete modes, are classified according to the mode numbers of the coupled structures, and this, among other things, depends on the number of periods of the device.
In a stellarator, a necessary condition for mode coupling is that the toroidal mode numbers $n_1$ and $n_2$ of two different modes differ by a multiple of the number of periods ($n_1\pm n_2=kN_{fp}$).
Modes that verify this condition belong to the same mode family, which is labeled according to the lower mode number of its constituents.
For instance, modes with \(n/m\) and $n/(m\pm 1)$ couple in the TAE gap and modes with \(n/m\) and $(n\pm\nu N_{fp})/(m\pm\mu)$ couple in the HAE$_{\mu\nu}$ gap.
TJ-II stellarator has four periods and this results in three distinct mode families ($N_f=0,1,2$).
These characteristics, specific to non-axisymmetric configurations, make the validation task particularly challenging.
References \cite{schwabIdealMagnetohydrodynamics1993, spongShearAlfven2003, kolesnichenkoAlfvenEigenmodes2002} provide a more in-depth treatment of the subject.
For the aforementioned reasons, a robust measurement of mode numbers and radial profile of the perturbation, combined with an accurate determination of the rotational transform profile, are both essential for theory validation purposes.

However, uncertainties in the mode number measurements, some due to the inherent limitations of the diagnostic and others that could be mitigated with an improved analysis technique, such as the SSI method [18] or Dynamic Mode Decomposition (DMD) [33], have been made clear in everything discussed in section \ref{sec:modeid}.
For instance, regarding inherent limitations to the mode number measurements, previous studies with a synthetic Mirnov diagnostic \cite{pons-villalongaExploringOperational2024} show that, besides uncertainties in poloidal mode number determination up to $\Delta m=2$, Mirnov coils arrays in TJ-II cannot distinguish different constituents in toroidally or helically coupled structures and therefore wherever theory predicts the destabilization of a coupled mode structure within a TAE or a HAE$_{\mu\nu}$ gap, it is very likely that only the numbers of the dominant mode or a combination of the numbers of both coupled modes can be measured reliably.
Another important inherent limitation, also discussed in \cite{pons-villalongaExploringOperational2024}, and common to all magnetic coil systems, is that the mode numbers of perturbations localized in the inner plasma can hardly be determined, such as, for example, the 220 kHz mode shown in figure \ref{fig:HIBP_mode_profiles_53493_90}.

A more accurate estimate of the evolution of the rotational transform profile could also be obtained by solving the current diffusion equation in the real TJ-II geometry \cite{strandMagneticFlux2001,schmittModelingMeasurement2010}.
The \iotabar\ evolution model presented in section \ref{sec:iotaevol} assumes that both the ECCD and NBCD currents are established instantaneously at the time at which the ECRH or NBI beams are injected.
This is a reasonable approximation, even for the case of NBI in which the current source needs a time of the order of several miliseconds to form.
Note that the slowing-down time of the fast ions ($\tau_{SD}$), which is actually longer for the present plasma parameters ($\tau_{SD}\approx 40$ ms from ASCOT5 simulations), is not a good proxy if we want to gain an insight into the time it takes for the NBCD current source to settle.
Taking $\tau_{SD}$ is actually misleading since the dominant contribution to the fast ion current does not come from thermalized ions but from the fast ones.
In fact, the lower the contents of fast ion in the plasma, the lower the current \cite{mulasValidatingNeutralbeam2023}.

As an additional uncertainty, the evolution of plasma profiles in the intermediate phase between the shutdown of ECRH2 and the entry of NBI2 is not considered.
A self-consistent calculation addressing the impact of NBCD on \iotabar\ would involve the use of a transport code such as \texttt{ASTRA} coupled to \texttt{ASCOT5}.
Despite the approximations, the \iotabar\ evolution model provides fairly reasonable results in terms of low-frequency modes, in a range in which the mode number analysis is robust (single n/m pair edge modes).

That the determination of the rotational transform and a robust measurement of mode numbers are essential for theory validation can be demonstrated by calculating the shear Alfvén continuum with the \texttt{STELLGAP} code \cite{spongShearAlfven2003} and observing the difficulties involved in identifying the modes that are destabilized in the experiment.
\texttt{STELLGAP} solves the Alfvén continuum equation for 3D stellarator equilibria taking into account the interactions between multiple toroidal modes.
The sign criterion for the phase term in \texttt{STELLGAP} ($m\theta-n\phi$) is opposite to the one used in the Lomb periodogram analysis and therefore $n$ must be replaced by $-n$ when comparing the results of the code to the measurements.

\begin{figure}[h]
	\centering
	\includegraphics[width=\linewidth]{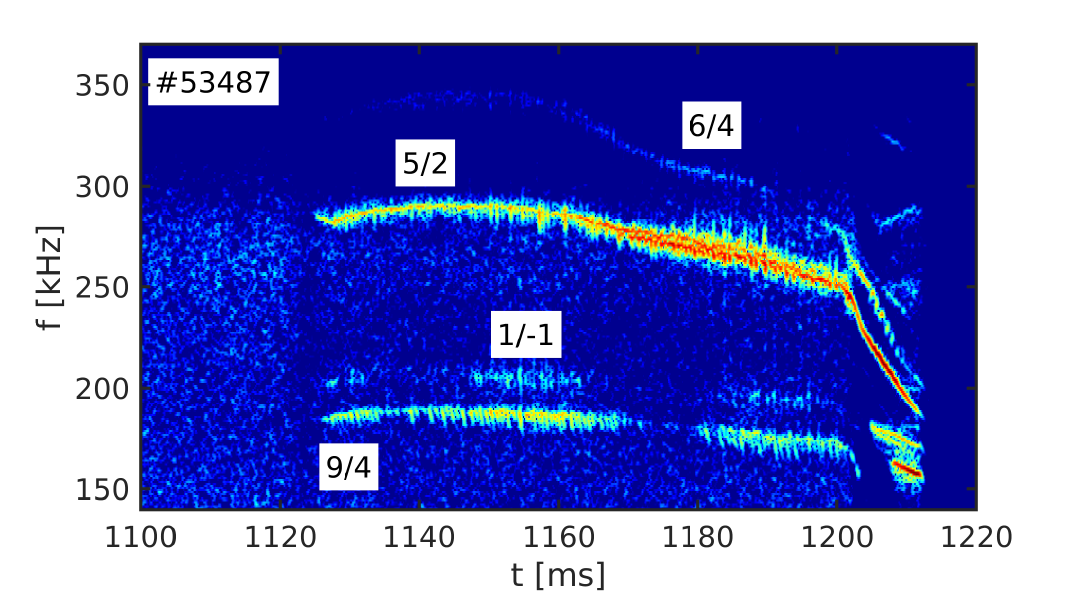}
	\includegraphics[width=\linewidth]{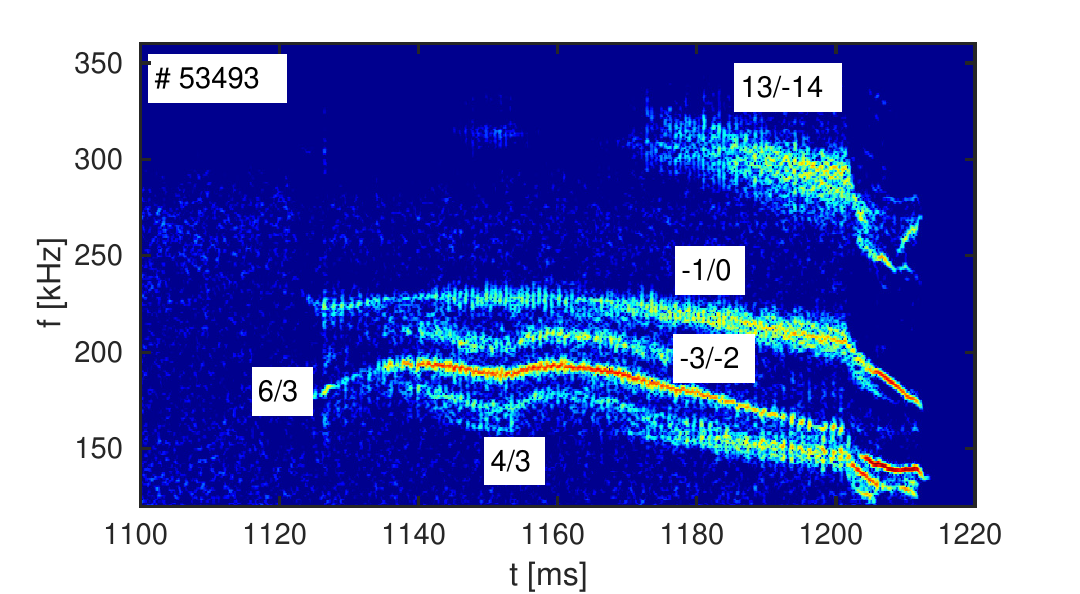}\\
	\caption{Spectrogram of magnetic fluctuations and measured mode numbers $n/m$ in medium \iotabar configuration for $n_{\parallel}=0$ (\#53487) and $n_{\parallel}=-0.1$ (\#53493).}
	\label{fig:zoom_modes_npar}
\end{figure}

\begin{figure}[h]
	\centering
	\includegraphics[width=0.75\linewidth]{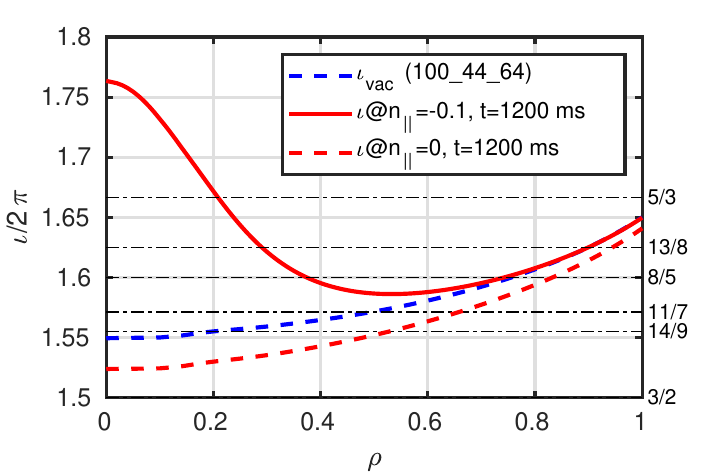}
	\caption{Rotational transform profiles of the medium \iotabar configuration at $t=1200$ ms for $n_{\parallel}=0$ (dashed red line) and $n_{\parallel}=-0.1$ (solid red line).
		Vaccum \iotabar\ is indicated by the dashed blue line.}
	\label{fig:stell_iotas}
\end{figure}

We have only analyzed a couple of cases to make our point, the ones belonging to the medium \iotabar\ configuration that exhibit the most diverse activity, i.e. $n_{\parallel}=0$ and $n_{\parallel}=-0.1$, and in which a clear change can be observed due to the appearance of a minimum in the \iotabar profile.
The spectrograms and measured mode numbers of each mode are shown in figure \ref{fig:zoom_modes_npar}.
The corresponding rotational transform profiles calculated at $t=1200$ ms, are shown in figure \ref{fig:stell_iotas}.

\begin{figure}[h]
	\centering
	\includegraphics[width=0.8\linewidth]{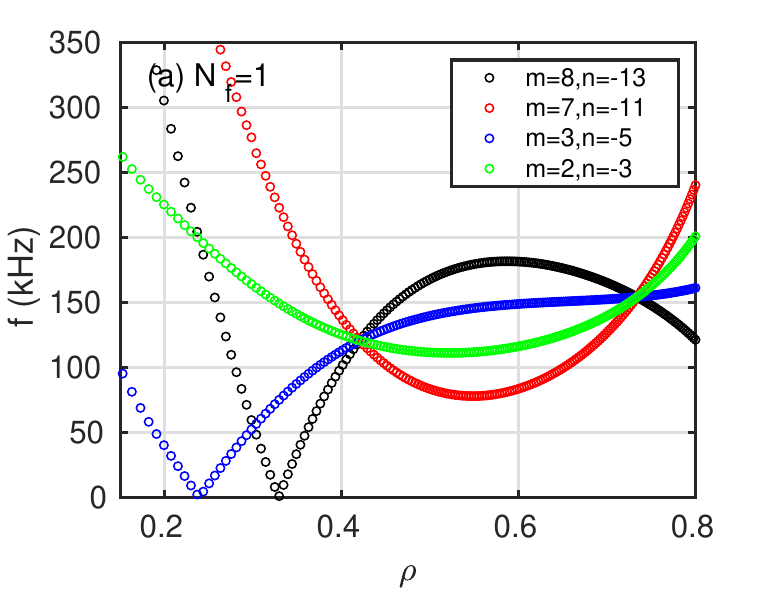}\\
	\includegraphics[width=0.8\linewidth]{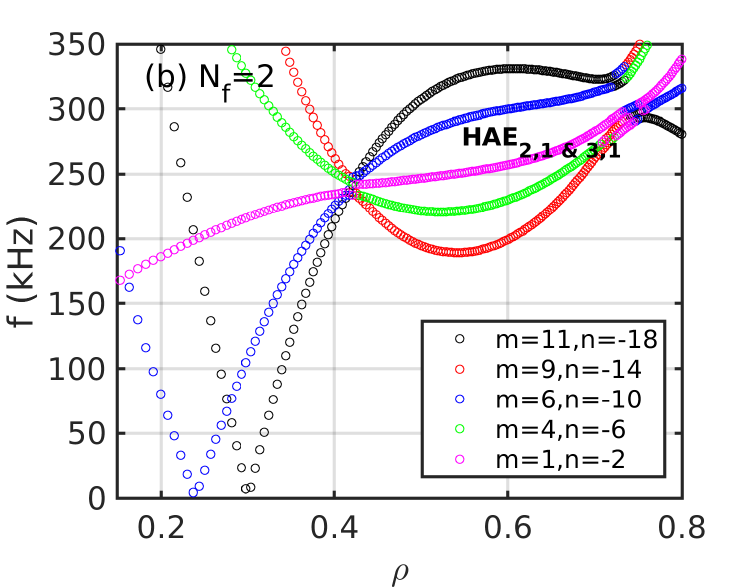}\\
	\caption{Shear Alfvén continuum for \(n_\parallel = -0.1\).}
	\label{fig:stell_mediota_npar-0.1}
\end{figure}

\begin{figure}[t]
	\centering
	\includegraphics[width=0.75\linewidth]{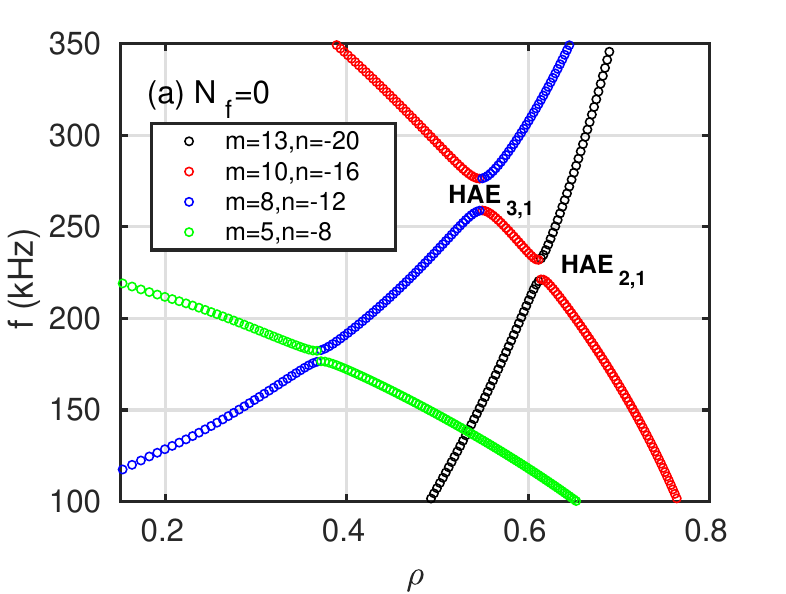}\\
	\includegraphics[width=0.75\linewidth]{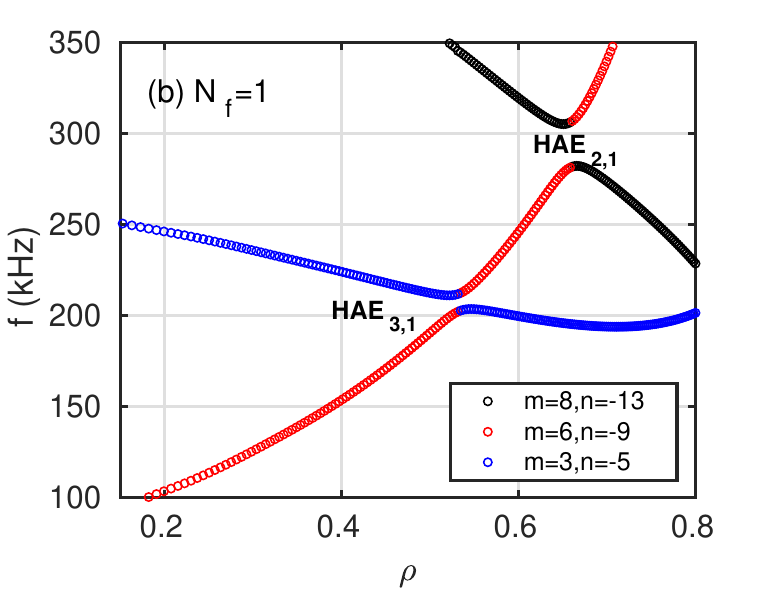}\\
	\includegraphics[width=0.75\linewidth]{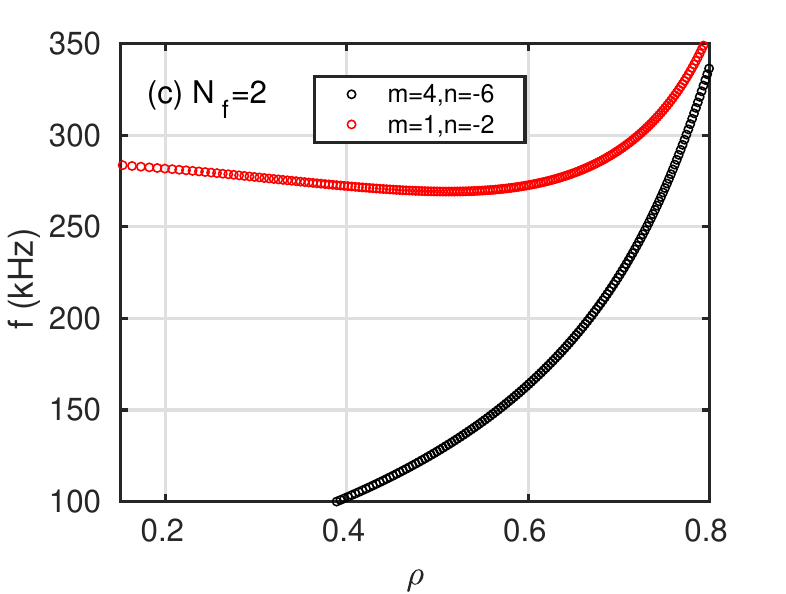}\\
	\caption{Shear Alfvén continuum for $n_{\parallel}=0$.
		For each family, only the relevant modes have been plotted.
		The position in frequency of the helical gaps is indicated.}
	\label{fig:stell_mediota_npar0}
\end{figure}

Figure \ref{fig:stell_mediota_npar-0.1} shows the result of the \texttt{STELLGAP} calculation for $n_{\parallel}=-0.1$ while figure \ref{fig:stell_mediota_npar0} shows the continuum for $n_{\parallel}=0$.
When $n_{\parallel}=-0.1$, a minimum in the rotational transform profile around $\rho\approx 0.55$ is produced by the combination of positive ECCD and negative NBCD.
Each plot corresponds to one mode family.
The low $\beta$ incompressible limit is considered here and coupling to sound waves is not included.
Moreover, no plasma impurities data has been used and $Z_{eff}=1$ is taken.
The minimum in \iotabar\ has a strong impact of the shear Alfvén spectrum as we could already guess from the experimental observation that shows up to five different modes in the magnetic fluctuations spectrum (see figure \ref{fig:zoom_modes_npar}).
The \iotabar\ minimum is responsible for the creation of maximums and minimums in frequency and favors the presence of global Alfvén Eigenmodes (GAEs).
The experimental frequency range covered by these modes goes from 150 to 350 kHz while the \texttt{STELLGAP} calculation spans from 75 to 325 kHz in the region around minimum \iotabar.

A direct comparison of measured mode numbers with the ones calculated using \texttt{STELLGAP} is subject to several uncertainties.
The first, most important, and most obvious, is that a calculation of the continuum only tells us which modes are possible in a given magnetic equilibrium.
Actually, estimates of destabilization rates by means of linear or non linear stability simulations using gyro-fluid or gyrokinetic codes such as \texttt{FAR3d} or \texttt{EUTERPE} are needed to provide an informed answer.
This also requires previous NBI simulations with ASCOT5 to determine the fast ion pressure profile and its energy distribution function.
Even so, beyond the clear impact that the minimum in \iotabar\ has both in experiments and theory, we can relate the observations to the calculated continuum.

For the $n_{\parallel}=-0.1$ case, the observed modes with low mode numbers $4/3$, $6/3$, very close in frequency, and determined with a high degree of reproducibility in the periodogram analysis (see again figure \ref{fig:lomb_results_aggregate}), are consistent with the calculated continuum of the $N_f=1$ family shown in figure \ref{fig:stell_mediota_npar-0.1}.
There, maximums and minimums in the continuum, also close in frequency, and prone to GAE destabilization appear around $\rho=0.55$ with mode numbers $-3/2$ and $-5/3$.
Differences between measured and calculated mode numbers would be easily explained by uncertainties in the mode number measurements.
The fact that we are neglecting sound coupling and we are using $Z_{eff}=1$ could also justify the differences in frequencies, which appear around 50 kHz lower in the calculation.
As for the $-1/0$ mode around 220 kHz, as we already mentioned above, HIBP measurements shows a radial structure localized in the inner plasma region, where mode number determination using magnetics is actually not reliable.
Finally, a high frequency $13/-14$ mode is observed around 310 kHz, where the calculation for \(N_f = 2\) predicts also high mode numbers AEs in the HAE$_{21}$ and HAE$_{31}$ gaps that merge in the plasma periphery.
HIBP measurements for some of these modes are shown in figure \ref{fig:HIBP_mode_profiles_53493_90}.

In the perpendicular injection case, for which no HIBP measurements were available, the comparison with the continuum is subject to many more uncertainties.
Two helical gaps (HAE$_{21}$ and HAE$_{31}$) in the spectrum appear for mode families $N_f=0$ and $N_f=1$.
The TAE gap appears higher in frequency and is not shown here.
According to the result shown in figure \ref{fig:zoom_modes_npar}, a $9/4$ mode is robustly identified around 175 kHz.
Most of the calculated gaps appear above this frequency (and we are actually undertestimating the calculated frequency) so it may not be an HAE mode.
The closest modes in frequency are the $\{-9/6$, $-5/3\}$ pair in the HAE$_{31}$ gap ($f\approx 200$ kHz) and the $\{-8/5, -12/8\}$ pair in the very narrow HAE$_{21}$ gap ($f\approx 175$ kHz) and it is difficult to know whether the $9/4$ mode is the footprint on the detection system of any of these combinations of modes without HIBP data and stability simulations.
The result of the experimental mode number analysis is not as conclusive when applied to the rest of observed modes.
A candidate to explain the destabilization of the brighter $5/2$ mode, that appear around 270 kHz, could be the $-5/3$ mode inside the HAE$_{31}$ gap.
Other modes, identified with less degree of confidence as $6/4$ and $1/-1$ could have its correspondence in the $N_f=2$ family.
The minimum in frequency around $\rho=0.5$ of the $-2/1$ mode favors the presence of a global Alfvén eigenmode (GAE) while the $6/4$ mode destabilized at $f\approx 300$ kHz mode could be an energetic particle mode (EPM) since it would intersect the continuum.
None of this can be assured with certainty without performing stability simulations using codes such as \texttt{FAR3d}, \texttt{EUTERPE} or \texttt{GTC}.

\section{Conclusions}

We have described a series of experiments performed in the TJ-II stellarator.
ECCD was used to modify the \iotabar\ profile in order to destabilize different NBI-driven Alfvén eigenmodes and to have a wide range of cases available for further validation studies.
The observed modes have been diagnosed with two arrays of Mirnov coils, used to determine the toroidal and poloidal mode numbers, and one heavy ion beam probe operated in radial sweep mode to measure the radial profile of the perturbation.
This is the first time that the toroidal mode number of Alvén waves in TJ-II has been measured.
Significantly, radial asymmetries were observed in the potential profiles of some modes, which points to coupled mode structures.

It is shown that changes in \iotabar\ that can lead to significant variations in the Alfvén continuum are not measurable with MSE and, instead, a cylindrical model for the evolution of the shielding current in response to the NBI and EC-induced currents has been used to estimate it.
The \iotabar\ profiles obtained with this model have been used as inputs for \texttt{STELLGAP}.
The calculation of the Alfvén continuum and its comparison with experimental results illustrates the complexity of this type of analysis in non-axisymmetric devices.
A couple of cases have been addressed and, within the uncertainties due both to the limitation of the Mirnov coils diagnostics and to the errors when estimating the rotational transform profile, the calculated results show a reasonable consistency between the shear Alfvén continua and the experimental observations.
Only the cases with the most intense activity have been analyzed with \texttt{STELLGAP}, and other interesting cases, such as the one exhibiting a structure with three very close frequencies and a clearly asymmetric radial profile, all compatible with time evolution of coupled modes, have been left for a more theoretical treatment.

The database of AEs obtained in different scenarios, characterized by their mode numbers and radial profiles of electrostatic potential can be used for theory validation purposes using gyro-fluid codes such as \texttt{FAR3d} \cite{varelaStabilityOptimization2024}, already applied to co-NBI driven modes in TJ-II \cite{cappaStabilityAnalysis2021} or gyrokinetic codes such as \texttt{EUTERPE} \cite{kleiberEUTERPEGlobal2024}, that was applied to study AEs fast-ion drive in W7-X \cite{slabyInvestigationMode2020}.
Thermal plasma profiles and estimates of neutral density profiles are available and fast ion slowing down simulations with \texttt{ASCOT5} in the three magnetic configurations can be performed to obtain the corresponding fast ion density profiles.
In addition to mode potential, HIBP measurements can also provide the density fluctuations profile associated to the Alfvén mode \cite{melnikov2DDistributions2022}, which are very relevant when comparing with the theoretical predictions.

\section{Acknowledgements}
The authors want to thank the TJ-II Team for their support.
This work was partially funded by the Spanish Ministry of Science and Innovation under several contracts, Grant Nos. ENE2013-48109-P, FIS2017-88892-P, FIS2017-85252-R, PID2021-125607NB-I00, and by the European Regional Development Fund (ERDF) “A way of making Europe”.
This work has been carried out within the framework of the EUROfusion Consortium, funded by the European Union via the Euratom Research and Training Programme (Grant Agreement No 101052200 — EUROfusion).
Views and opinions expressed are however those of the author(s) only and do not necessarily reflect those of the European Union or the European Commission.
Neither the European Union nor the European Commission can be held responsible for them.

\FloatBarrier

\appendix
\section{Radial diffusion of toroidal current}
\label{app:rad_diffusion}

We start from Maxwell's equations where we neglect the displacement current term,
\begin{align}
	\nabla\times\mathbf{E} = & -\frac{\partial\mathbf{B}}{\partial t} \label{eq:max1} \\
	\nabla\times\mathbf{B} = & \mu_0 \mathbf{J},
	\label{eq:max2}
\end{align}
and we assume that the toroidal current density $\mathbf{J}(r,t)$ is related to the electric field through Ohm's law,
\begin{equation}
	\mathbf{J}(r,t)=\mathbf{J_s}(r)+\sigma(r)\mathbf{E}(r,t).
	\label{eq:current_dens}
\end{equation}
where $\mathbf{J_s}(r)$ and $\mathbf{J_E}(r,t)\equiv\sigma(r)\mathbf{E}(r,t) $ are the current densities of the source and shielding terms respectively.
Combining equations \ref{eq:max1} and \ref{eq:max2}, assuming only radial dependencies of all quantities
and assuming also vector quantities with components only along the direction parallel to the static field $\mathbf{B_0}$, we get the diffusion equation for the shielding current density in cylindrical coordinates
\begin{equation}
	\mu_0\frac{\partial J_E(r,t)}{\partial t}=\nabla^2 E(r,t)
\end{equation}
which can be written as
\begin{equation}
	\mu_0\frac{\partial J_E(r,t)}{\partial t}=\frac{1}{r}\left[\frac{\partial}{\partial r}\left(r\frac{\partial}{\partial r} \left[\eta(r) J_E(r,t)\right] \right) \right],
	\label{eq:scurrentd_diff}
\end{equation}
where $\eta(r)=1/\sigma(r)$ is the plasma resistivity. Using the relation between current density and integrated current
\begin{equation}
	I_E(r,t)=2\pi\int_{0}^{r} J_E(r^\prime,t) r^\prime dr^\prime.
	\label{eq:int_current_dens}
\end{equation}
and integrating equation \ref{eq:scurrentd_diff} we obtain
\begin{equation}
	\frac{\mu_0}{2\pi}\frac{\partial I_E(r,t)}{\partial t}=r\frac{\partial}{\partial r} \left[\eta(r) J_E(r,t)\right].
	\label{eq:scurrent_diff}
\end{equation}
After some algebraic manipulations and noting that $2\pi r J_E=\partial I_E/\partial r$, we arrive at the following equation for the time evolution of the shielding current \(I_E(r,t)\),
\begin{equation}
	\mu_0\frac{\partial I_E(r,t)}{\partial t}=r\left[\frac{\partial}{\partial r}\left(\frac{\eta(r)}{r}\frac{\partial I_E(r,t)}{\partial r}\right) \right].
	\label{eq:radialdif}
\end{equation}
Equation \ref{eq:radialdif} is equivalent to the one given in \cite{zaniniECCDinducedSawtooth2020} taking into account that, in the cylindrical case, the relation between any current component and the
variation that it induces in iota is given by
\begin{equation}
	\iotabar(r,t)=\frac{\mu_0 R}{2\pi B_0}\frac{I(r,t)}{r^2}.
\end{equation}
For a cylindrical plasma with minor radius $a$, central resistivity $\eta_0$ and time constant $\tau_{LR}\equiv\mu_0 a^2/\eta_0$, we can introduce dimensionless variables $\tau\equiv t/\tau_{LR}$, $\rho\equiv r/a$ and $\eta^*\equiv\eta/\eta_0$ and write equation \ref{eq:radialdif} for $I_E(\rho,\tau)$ as
\begin{equation}
	\frac{1}{\eta^*}\frac{\partial I_E}{\partial \tau}=\frac{\partial}{\partial \rho}\left(\frac{\partial I_E}{\partial \rho}\right)+N(\rho)\frac{\partial I_E}{\partial \rho},
	\label{eq:adim_radialdif}
\end{equation}
where we have defined $N(\rho)\equiv (\partial\eta^*/\partial\rho)/\eta^*-1/\rho$.
We solve this equation numerically with a standard \texttt{MATLAB}\,\textsuperscript{\tiny\textregistered} solver.
Once $I_E(\rho,\tau)$ is known we may write the total rotational transform as
\begin{align}
	\iotabar(\rho,\tau)= & \iotabar_{vac}(\rho)+\iotabar_s(\rho)+\iotabar_E(\rho,\tau)          \label{eq:iota_evol} \\
	=                    & \iotabar_{vac}(\rho)+\frac{\mu_0 R}{2\pi B_0a^2\rho^2}(I_s(\rho)+I_E(\rho,\tau))
\end{align}

\pagebreak

\section{Mode number results}
\label{app:mode_tables}

This appendix contains tables that summarize the mode number analysis.
In them, only the modes that appear as most intense more than 15\% of the time are shown, along with the total number of analyzed time intervals.
If no mode appears the minimum amount of times, then no mode numbers are shown.

\begin{table}[h]
	\centering
	\begin{tabular}{@{}lrcr@{}}
		\toprule
		\(n_\parallel\) & f [kHz] & (\(n, m\)) & Counts    \\ \midrule
		-0.2            & 260.0   & ---        & ---       \\ \cmidrule(l){4-4}
		                &         &            & Total: 22 \\ \midrule
		-0.1            & 17.4    & (-3, -2)   & 14        \\
		                &         & (1, -1)    & 5         \\ \cmidrule(l){4-4}
		                &         &            & Total: 21 \\ \midrule
		-0.1            & 368.2   & (13, -15)  & 8         \\ \cmidrule(l){4-4}
		                &         &            & Total: 45 \\ \midrule
		0.0             & 13.0    & (-3, -2)   & 12        \\ \cmidrule(l){4-4}
		                &         &            & Total: 12 \\ \midrule
		0.0             & 273.2   & (-14, -3)  & 29        \\ \cmidrule(l){4-4}
		                &         &            & Total: 33 \\ \midrule
		0.1             & 12.7    & (-3, -2)   & 10        \\ \cmidrule(l){4-4}
		                &         &            & Total: 14 \\ \midrule
		0.1             & 251.9   & (6, 2)     & 12        \\
		                &         & (13, 5)    & 6         \\ \cmidrule(l){4-4}
		                &         &            & Total: 30 \\ \midrule
		0.2             & 11.0    & (-3, -2)   & 12        \\ \cmidrule(l){4-4}
		                &         &            & Total: 14 \\ \bottomrule
	\end{tabular}
	\caption{Counts of mode numbers (\(n,m\)) at different frequencies for the low iota (100\_36\_62) configuration.}
	\label{tab:mode_counts_lowiota}
\end{table}

\begin{table}[h]
	\centering
	\begin{tabular}{@{}lrcr@{}}
		\toprule
		\(n_\parallel\) & f [kHz] & (\(n, m\)) & Counts    \\ \midrule
		-0.2            & 47.1    & (-9, -5)   & 13        \\
		                &         & (-8, -5)   & 3         \\ \cmidrule(l){4-4}
		                &         &            & Total: 18 \\ \midrule
		-0.2            & 219.8   & (-14, -6)  & 10        \\ \cmidrule(l){4-4}
		                &         &            & Total: 32 \\ \midrule
		-0.2            & 153.6   & (-9, 1)    & 5         \\
		                &         & (11, 4)    & 5         \\ \cmidrule(l){4-4}
		                &         &            & Total: 30 \\ \midrule
		-0.2            & 301.8   & (-1, 3)    & 8         \\
		                &         & (13, -14)  & 7         \\ \cmidrule(l){4-4}
		                &         &            & Total: 32 \\ \midrule
		-0.1            & 17.3    & (-9, -5)   & 11        \\ \cmidrule(l){4-4}
		                &         &            & Total: 12 \\ \midrule
		-0.1            & 187.2   & (6, 3)     & 13        \\ \cmidrule(l){4-4}
		                &         &            & Total: 15 \\ \midrule
		-0.1            & 221.0   & (-1, 0)    & 5         \\
		                &         & (0, 0)     & 4         \\
		                &         & (-2, 0)    & 2         \\ \cmidrule(l){4-4}
		                &         &            & Total: 13 \\ \midrule
		-0.1            & 206.6   & (-3, -2)   & 4         \\
		                &         & (1, -1)    & 4         \\ \cmidrule(l){4-4}
		                &         &            & Total: 8  \\ \midrule
		-0.1            & 165.0   & (4, 3)     & 7         \\ \cmidrule(l){4-4}
		                &         &            & Total: 13 \\ \midrule
		-0.1            & 301.6   & (13, -14)  & 5         \\ \cmidrule(l){4-4}
		                &         &            & Total: 9  \\ \bottomrule
	\end{tabular}
	\caption{Counts of mode numbers (\(n,m\)) at different frequencies for the standard iota (100\_44\_64) configuration.}
	\label{tab:mode_counts_mediota_1}
\end{table}

\begin{table}[h]
	\centering
	\begin{tabular}{@{}lrcr@{}}
		\toprule
		\(n_\parallel\) & f [kHz] & (\(n, m\)) & Counts    \\ \midrule
		0.0             & 20.3    & (-9, -5)   & 28        \\ \cmidrule(l){4-4}
		                &         &            & Total: 31 \\ \midrule
		0.0             & 261.1   & (5, 2)     & 11        \\ \cmidrule(l){4-4}
		                &         &            & Total: 31 \\ \midrule
		0.0             & 177.8   & (9, 4)     & 22        \\
		                &         & (-3, 1)    & 15        \\ \cmidrule(l){4-4}
		                &         &            & Total: 38 \\ \midrule
		0.0             & 193.7   & (1, -1)    & 11        \\
		                &         & (-3, 0)    & 5         \\ \cmidrule(l){4-4}
		                &         &            & Total: 31 \\ \midrule
		0.0             & 294.4   & (-12, -1)  & 10        \\
		                &         & (6, 4)     & 8         \\ \cmidrule(l){4-4}
		                &         &            & Total: 30 \\
		0.1             & 21.1    & (-9, -5)   & 10        \\
		                &         & (-8, -5)   & 4         \\ \cmidrule(l){4-4}
		                &         &            & Total: 21 \\ \midrule
		0.1             & 275.3   & (12, 5)    & 9         \\
		                &         & (0, 5)     & 5         \\
		                &         & (-7, -1)   & 5         \\ \cmidrule(l){4-4}
		                &         &            & Total: 32 \\ \midrule
		0.2             & 21.6    & (-9, -5)   & 8         \\
		                &         & (-8, -5)   & 5         \\ \cmidrule(l){4-4}
		                &         &            & Total: 14 \\ \midrule
		0.2             & 270.8   & (12, 5)    & 10        \\
		                &         & (13, 5)    & 7         \\ \cmidrule(l){4-4}
		                &         &            & Total: 19 \\ \bottomrule
	\end{tabular}
	\caption{Counts of mode numbers (\(n,m\)) at different frequencies for the standard iota (100\_44\_64) configuration (cont.).}
	\label{tab:mode_counts_mediota_2}
\end{table}

\begin{table}[h]
	\centering
	\begin{tabular}{@{}lrcr@{}}
		\toprule
		\(n_\parallel\) & f [kHz] & (\(n, m\)) & Counts    \\
		\midrule
		-0.2            & 38.5    & (14, 0)    & 4         \\ \cmidrule(l){4-4}
		                &         &            & Total: 21 \\ \midrule
		-0.2            & 192.8   & (-10, -1)  & 7         \\ \cmidrule(l){4-4}
		                &         &            & Total: 41 \\ \midrule
		-0.2            & 271.1   & ---        & ---       \\ \cmidrule(l){4-4}
		                &         &            & Total: 24 \\ \midrule
		-0.1            & 55.6    & (-9, -5)   & 14        \\
		                &         & (-8, -5)   & 3         \\ \cmidrule(l){4-4}
		                &         &            & Total: 19 \\ \midrule
		-0.1            & 191.1   & (-15, -5)  & 13        \\ \cmidrule(l){4-4}
		                &         &            & Total: 39 \\ \midrule
		-0.1            & 204.1   & (6, 1)     & 12        \\
		                &         & (7, 1)     & 8         \\ \cmidrule(l){4-4}
		                &         &            & Total: 28 \\ \midrule
		-0.1            & 287.6   & (2, 2)     & 5         \\ \cmidrule(l){4-4}
		                &         &            & Total: 21 \\
		0.0             & 27.3    & (-14, -8)  & 10        \\
		                &         & (11, -2)   & 6         \\ \cmidrule(l){4-4}
		                &         &            & Total: 26 \\ \midrule
		0.0             & 52.1    & (-9, -5)   & 23        \\ \cmidrule(l){4-4}
		                &         &            & Total: 28 \\ \midrule
		0.0             & 195.4   & (-14, -6)  & 21        \\
		                &         & (-3, 2)    & 6         \\ \cmidrule(l){4-4}
		                &         &            & Total: 34 \\ \midrule
		0.0             & 247.1   & (1, 0)     & 3         \\ \cmidrule(l){4-4}
		                &         &            & Total: 20 \\ \bottomrule
	\end{tabular}
	\caption{Counts of mode numbers (\(n,m\)) at different frequencies for the high iota (100\_49\_65) configuration.}
	\label{tab:mode_counts_highiota_1}
\end{table}

\begin{table}[h]
	\centering
	\begin{tabular}{@{}lrcr@{}}
		\toprule
		\(n_\parallel\) & f [kHz] & (\(n, m\)) & Counts    \\
		\midrule
		0.1             & 27.6    & (-14, -8)  & 6         \\
		                &         & (-1, -5)   & 5         \\ \cmidrule(l){4-4}
		                &         &            & Total: 25 \\ \midrule
		0.1             & 49.4    & (-9, -5)   & 16        \\ \cmidrule(l){4-4}
		                &         &            & Total: 18 \\ \midrule
		0.1             & 251.8   & (11, 6)    & 12        \\
		                &         & (-15, -3)  & 7         \\ \cmidrule(l){4-4}
		                &         &            & Total: 32 \\ \midrule
		0.1             & 188     & (1, 0)     & 7         \\ \cmidrule(l){4-4}
		                &         &            & Total: 23 \\ \midrule
		0.2             & 28.4    & (-9, -5)   & 5         \\
		                &         & (-1, -5)   & 4         \\ \cmidrule(l){4-4}
		                &         &            & Total: 21 \\ \midrule
		0.2             & 55      & (-9, -5)   & 11        \\ \cmidrule(l){4-4}
		                &         &            & Total: 17 \\ \midrule
		0.2             & 233.6   & (5, 2)     & 19        \\
		                &         & (-7, -1)   & 17        \\ \cmidrule(l){4-4}
		                &         &            & Total: 36 \\ \midrule
		0.2             & 181.7   & (14, 3)    & 7         \\
		                &         & (1, 0)     & 6         \\
		                &         & (2, 0)     & 3         \\ \cmidrule(l){4-4}
		                &         &            & Total: 20 \\ \bottomrule
	\end{tabular}
	\caption{Counts of mode numbers (\(n,m\)) at different frequencies for the high iota (100\_49\_65) configuration (cont.).}
	\label{tab:mode_counts_highiota_2}
\end{table}

\FloatBarrier

\printbibliography

\end{document}